\documentclass[
  a4paper,
  UKenglish,
  cleveref,
]{lipics-v2021}
\usepackage{siunitx}
\usepackage{mathtools,bm}
\usepackage{tabularray}
\usepackage{tikz}

\usetikzlibrary{arrows.meta,positioning,calc,fit}

\UseTblrLibrary{booktabs}

\newcommand{\etal}{\emph{et al.}}

\newcommand{\eg}{\emph{e.g.}}
\newcommand{\ie}{\emph{i.e.}}

\crefname{observation}{observation}{observations}
\Crefname{observation}{Observation}{Observations}
\hideLIPIcs
\nolinenumbers

\title{%
Which \texorpdfstring{\(\Diamond\mathcal{P}\)}{Eventually Perfect} Algorithm is Suitable for You?
An Empirical Benchmark on Real Network Traces%
}
\titlerunning{Which \texorpdfstring{\(\Diamond\mathcal{P}\)}{Eventually Perfect} Algorithm is Suitable?}
\author{Yenan Wang}{Chalmers University of Technology, Sweden}{yenan@chalmers.se}{https://orcid.org/0009-0004-1389-1196}{}
\author{Jens Kollberg}{Chalmers University of Technology, Sweden}{jens.collberg01@gmail.com}{}{}
\author{Elad Michael Schiller}{Chalmers University of Technology, Sweden}{elad@chalmers.se}{https://orcid.org/0000-0003-3258-3696}{}
\author{Carla Fabiana Chiasserini}{ Chalmers University of Technology, Sweden \and Politecnico di Torino, Italy}{carla.chiasserini@polito.it}{https://orcid.org/0000-0003-1410-660X}{}
\authorrunning{Y. Wang, J. Kollberg, E. M. Schiller, and C. F. Chiasserini}
\Copyright{Yenan Wang, Jens Kollberg, Elad Michael Schiller, and Carla Fabiana Chiasserini}

\ccsdesc[500]{Computing methodologies~Distributed algorithms}
\ccsdesc[500]{Computer systems organization~Dependable and fault-tolerant systems and networks}

\keywords{%
Empirical evaluation,
\texorpdfstring{\(\Diamond\mathcal{P}\)}{Diamond P} failure detectors,
practical recommendations%
}

\supplementdetails[subcategory={Source Code},cite=ufd-eval]{Software}{https://bitbucket.org/eladschiller/ufd-eval/}
\supplementdetails[subcategory={Network Traces},cite=dataset-2026-24465359]{Dataset}{https://doi.org/10.5281/zenodo.24465359}

\acknowledgements{We thank Ant\'onio Casimiro and M\^onica Dixit for providing us with the source code of Adaptare-FD~\cite{DBLP:conf/srds/DixitC10}, which allowed us to quickly integrate it into our evaluation.}

\begin{document}
\maketitle
\begin{abstract}

Unreliable failure detectors (UFDs) are a fundamental abstraction for dependability and fault tolerance in distributed systems, supporting core distributed algorithms, such as consensus.
Yet selecting a suitable detector remains difficult because existing implementations are evaluated under different environments, workloads, parameterizations, and quality-of-service (QoS) requirements.
We address this problem by systematizing 14 \(\Diamond \mathcal{P}\) and closely related \(\Diamond \mathcal{P}\)-style UFDs, implementing 13 of them, and providing a reusable benchmark for their empirical comparison.
Our methodology evaluates detectors through common QoS metrics, such as detection time, mistake rate, and query accuracy, and combines same-input replay of network traces with live deployment for detectors whose active probing or cooperative behavior requires online execution.
We then formalize failure-detector selection as a constrained algorithm-selection challenge, mapping operational requirements to suitable detector candidates.
We instantiate this methodology under controlled variations of communication rate, host load, and crash regime.
Applying the selection process shows that no detector is uniformly preferable across the evaluated settings.
At relaxed detection-time budgets, 2W~[Tomsic \etal, IPDPS 2015] is the most frequently selected detector, appearing in 12 of 16 scenario–metric combinations for \(T_D \le \qty{500}{\ms}\), whereas stricter budgets yield substantially more scenario- and metric-dependent selections.
The benchmark and selection methodology provide a reusable basis for practical detector selection and for establishing stronger analytical baselines for future failure-detection approaches, including learning-based detectors.

\end{abstract}

\clearpage
\section{Introduction}\label{sec:introduction}

Consensus is a conceptually simple task: a group of distributed nodes must agree on a single value.
Yet deterministic consensus cannot guarantee termination in a fully asynchronous system if even one node may crash~\cite{DBLP:journals/jacm/FischerLP85}.
The fundamental difficulty is that a node observing silence cannot determine whether another node has crashed or is merely slow.
Consensus therefore requires additional assumptions or information about the system.

One approach is to strengthen the model with timing assumptions, as in partial synchrony~\cite{DBLP:journals/jacm/DworkLS88}.
Chandra and Toueg instead separate such assumptions from the upper-layer algorithm through the unreliable failure detector (UFD) abstraction~\cite{DBLP:journals/jacm/ChandraT96}.
The distributed algorithm reasons using the UFD's output, while the UFD encapsulates the mechanisms and assumptions used to infer crashes.
A UFD may make mistakes provided that its outputs satisfy the completeness and accuracy properties required by its class.

Chandra and Toueg~\cite{DBLP:journals/jacm/ChandraT96} define several UFD classes.
In the standard crash-prone setting with a majority of correct processes, the eventual-leader abstraction $\Omega$ captures the weakest failure information needed for consensus~\cite{DBLP:journals/jacm/ChandraHT96}.
We instead focus on the stronger Eventually Perfect class, $\Diamond\mathcal{P}$, which satisfies \emph{strong completeness} (eventually every crashed process is permanently suspected by every correct process) and \emph{eventual strong accuracy} (after some finite time, no correct process is suspected by any correct process)~\cite{DBLP:journals/jacm/ChandraT96}.
This per-process trust/suspect behavior makes false suspicions and detection time directly meaningful operational quantities.
We therefore use $\Diamond\mathcal{P}$ as our conceptual anchor while also considering closely related detectors that exhibit $\Diamond\mathcal{P}$-style behavior, producing a trust/suspect output after convergence (\Cref{tab:fd_proof_status}).

\begin{table}[ht]
\centering
\caption{Relationship of the studied failure detectors to $\Diamond\mathcal{P}$. For compactness, we omit the suffix ``FD'' from most detector names.
This naming convention applies to ED-FD~\cite{EDFD/Xiong07}, Adaptare-FD~\cite{DBLP:conf/srds/DixitC10}, 2W-FD~\cite{DBLP:conf/ipps/TomsicSGAS15}, WD-FD~\cite{WeibullFD/journal.pone.0173666}, 2E-FD~\cite{2EFD/journal.pone.0191577}, Stab-FD, and Stab\(^C\)-FD~\cite{DBLP:journals/jpdc/SensARM24}.
B-AFD is systematized but not implemented; we evaluate instead its extension S-AFD (\Cref{sec:chronology}).
}
\label{tab:fd_proof_status}

\begin{tblr}{
    colspec = {
        Q[l, wd=0.4\textwidth]
        X[l]
    },
    colsep = 0.5em,
    rowsep = 0.15em,
    row{1} = {font=\bfseries, c},
    column{1} = {valign=m},
    column{2} = {valign=m},
    row{even} = {bg=gray!10},
}
\toprule
Status & Detectors \\
\midrule
Proven as $\Diamond\mathcal{P}$ detector
& Bertier~\cite{DBLP:conf/dsn/BertierMS02}, 2E~\cite{2EFD/journal.pone.0191577} \\

Accrual formulation with thresholdable $\Diamond\mathcal{P}$-style behavior
& $\phi$-FD~\cite{DBLP:conf/srds/HayashibaraDYK04, DBLP:conf/dsn/DefagoUHK05},
ED~\cite{EDFD/Xiong07,DBLP:conf/dsn/DefagoUHK05},
WD~\cite{WeibullFD/journal.pone.0173666,DBLP:conf/dsn/DefagoUHK05} \\

QoS analysis with binary trust/suspect behavior
& Chen~\cite{DBLP:journals/tc/ChenTA02a},
B-AFD~\cite{DBLP:journals/sensors/DongRZL14}
(extended by S-AFD~\cite{S-AFD/Liu15}) \\

Engineering-oriented $\Diamond\mathcal{P}$-style behavior
& Satzger~\cite{DBLP:conf/sac/SatzgerPTU07},
BDBD~\cite{DBLP:conf/netys/BlanchardDBD14},
Adaptare~\cite{DBLP:conf/srds/DixitC10},
S-AFD~\cite{S-AFD/Liu15},
2W~\cite{DBLP:conf/ipps/TomsicSGAS15},
Stab and Stab$^C$~\cite{DBLP:journals/jpdc/SensARM24} \\
\bottomrule
\end{tblr}
\end{table}

\subparagraph*{Scope.}
The term ``failure detector'' is broader in systems research than the Chandra–Toueg oracle abstraction studied here.
For example, Rapid~\cite{DBLP:conf/usenix/SureshMGCL18} combines multi-source failure evidence with consistent membership changes, uKharon~\cite{DBLP:conf/usenix/GuerraouiMPXYZ22} integrates a multi-level detector with RDMA-based membership and reconfiguration, and FiDe~\cite{DBLP:conf/usenix/RovelliCBPJE25} obtains highly reliable microsecond-scale crash reporting through datacenter-specific mechanisms.
These systems target interfaces and guarantees that differ from the UFD abstraction considered here.
Unless stated otherwise, we therefore use \emph{failure detector} to mean the UFDs within our scope.

The eventual guarantees of \(\Diamond \mathcal{P}\) are qualitative: they do not specify how quickly crashes are detected or how frequently correct processes are falsely suspected.
Chen~\etal~\cite{DBLP:journals/tc/ChenTA02a} introduced implementation-independent quality-of-service (QoS) metrics for these operational properties, allowing detectors to be compared directly through detection speed and accuracy metrics, without embedding each detector into a particular consensus protocol, whose design and assumptions would introduce additional effects.
We therefore adopt failure-detector QoS as our common evaluation interface, adapting the metrics to our empirical setting.

Learning-based UFD-style detection has also been explored, from neural-network heartbeat prediction~\cite{10.1007/failure-detectors-nn} to recent LSTM-based approaches~\cite{DBLP:journals/corr/abs-2210-00134}.
Such approaches introduce training, inference, and deployment costs and choices absent from conventional UFDs, and existing proposals do not establish that the learned detector satisfies a Chandra–Toueg failure-detector class~\cite{Qi2023/10.1117/12.3004136,Yao2025/10.1145/3725536.3725549}.
Meaningful evaluation of learning-based detection therefore requires comparison against strong UFD alternatives under the same operational conditions and QoS requirements.

\subparagraph*{Research Gaps and Questions.}
To our knowledge, no open-source framework implements and evaluates \(\Diamond\mathcal{P}\)/\(\Diamond\mathcal{P}\)-style detectors under a common methodology.
Existing evaluations are fragmented across implementations, traces, platforms, workloads, parameterizations, and comparison sets, and individual detectors are typically compared against only a small subset of prior work.
Consequently, published results cannot be combined into a reliable cross-detector ranking.
The same limitation affects ML evaluations: outperforming Chen~\cite{DBLP:journals/tc/ChenTA02a} or another individual baseline does not establish superiority over the best conventional UFD alternative for the same operating point.
Moreover, Chen-style QoS metrics characterize detector behavior, but do not tell practitioners which detector to select for a given operational environment.
An unsuitable detector may react too aggressively, causing unnecessary recovery or reconfiguration, or too conservatively, delaying fault recovery.
We therefore ask:
\begin{description}
    \item[RQ1:] How do the QoS trade-offs of the studied detectors change with network conditions, communication rate, host load, crash frequency, and detection-time requirements?
    \item[RQ2:] How can the observed trade-offs be mapped to suitable detector candidates under specified deployment and application constraints?
\end{description}

\subparagraph*{Approach.}
To address these research gaps, we provide a reusable benchmark for evaluating and selecting among $\Diamond\mathcal{P}$ and $\Diamond\mathcal{P}$-style failure detectors.
We systematize the relevant design space and implement 13 detectors under a common QoS evaluation framework.
Where possible, detectors are evaluated through offline replay of identical network traces; detectors requiring active probing or cooperation are evaluated through live deployment.

Building on this benchmark, we formalize failure-detector selection as a constrained instance of Rice's Algorithm Selection Problem~\cite{DBLP:journals/ac/Rice76}. 
Operational conditions characterize the instance, the studied detectors form the algorithm portfolio, and their measured QoS metrics characterize performance.
Given a maximum acceptable detection time, we apply an $\epsilon$-constraint formulation~\cite{Haimes1971_EpsilonConstraint} to minimize the specified accuracy loss among feasible detectors.

We use a rapid-deployment data-collection tool on the PlanetLab platform~\cite{PlanetLab/10.1145/956993.956995} to demonstrate the methodology and selection process under controlled variations of communication rate, host load, and crash regimes.
Our open-source framework and UFD implementations are designed to be reusable and can be deployed on other real-world distributed systems.

\subparagraph*{Findings.}
Our results show no one-size-fits-all choice across the evaluated operating conditions and QoS objectives.
Performance depends strongly on the required detection-time threshold and accuracy metric, while communication rate and observed execution conditions can further alter the attainable trade-offs (\Crefrange{obs:communication-rate-effect}{obs:online-detectors}, \Cref{sec:results}).
Nevertheless, some patterns emerge.
For example, at relaxed detection-time requirements (\eg, \(T_D \le 500, \qty{2000}{\ms}\)), 2W is the most frequently selected
detector, appearing in 12 of 16 and 10 of 16 evaluated scenario--metric combinations, respectively (\Cref{sec:applying_selection_method}).
Under stricter detection-time budgets, however, the selected detector depends more strongly on the evaluated scenario and target metrics.
These results further motivate our constraint-based selection procedure in \Cref{sec:selection_process}, which maps measured QoS trade-offs and application requirements to suitable detector candidates rather than assuming a universal ranking.

\subparagraph*{Related Work.}
Raynal~\cite{IntroToFD/10.1145/1052796.1052806} surveys failure detectors as modular oracles characterized by completeness and accuracy, while Chaurasia and Verma~\cite{StudyOnFD/Chaurasia2020} survey a broader range of failure models, system assumptions, and implementation strategies.
These works organize the literature but do not provide a controlled empirical benchmark for selecting among multiple $\Diamond\mathcal{P}$ or $\Diamond\mathcal{P}$-style implementations.

Empirical failure-detection studies include mobile ad-hoc networks~\cite{DBLP:journals/ijpcc/FriedmanT09}.
More recent work includes formal \(\Diamond \mathcal{P}\) implementations for specialized communication structures, such as ADD networks~\cite{DBLP:journals/ppl/KumarW19,DBLP:journals/ppl/VargasRR20,DBLP:conf/netys/RumreichS22} and vCube-based hierarchical monitoring~\cite{DBLP:conf/ladc/SteinRDA23}, while IHFD~\cite{Chaurasia2026IHFD} evaluates hierarchical health monitoring using PlanetLab traces.
These approaches make communication topology, channel assumptions, or health aggregation part of the detector design, introducing benchmark dimensions orthogonal to the QoS-oriented design choices studied here.
We therefore leave such topology- or architecture-coupled detection to extensions of the benchmark.

Prior studies with comparative evaluations~\cite{DBLP:conf/srds/HayashibaraDYK04,EDFD/Xiong07,DBLP:journals/jsac/XiongVYSPKL09} replay recorded traces to a few heartbeat-based detectors.
We extend this to 13 detectors, including pull-based, cooperative, and self-stabilizing designs.
Our work complements these efforts by providing a broader reusable implementation corpus, common-input evaluation where possible, and a formal methodology for selecting among detectors according to operational conditions and application requirements.

\subparagraph*{Contributions and Implications.}
Our contributions are the following:
\begin{enumerate}
\item We provide a reusable benchmarking framework for evaluating $\Diamond\mathcal{P}$/$\Diamond\mathcal{P}$-style failure detectors,  together with reusable artifacts for data collection, trace replay, and deployment.
We further systematize the relevant design space and provide implementations of 13 detectors spanning timeout-based, accrual, probe-based, cooperative, and self-stabilizing designs~(\Cref{sec:sok_of_fds}).

\item We present a common QoS evaluation methodology that combines same-input trace replay with live evaluation where active interaction is intrinsic to the detector~(\Crefrange{sec:methods}{sec:evaluation}).

We then demonstrate its rapid deployment through a controlled comparison of these detectors on real-world testbeds.
We characterize detector performance with communication rate, host workload, crash failures, and detection-time requirements, revealing distinct performance regimes rather than a universally preferable detector~(\Cref{sec:results}).

\item We formalize failure-detector selection as a constrained algorithm-selection challenge, building on Rice's Algorithm Selection Problem~\cite{DBLP:journals/ac/Rice76}.
This provides a principled basis for mapping operational conditions and empirical QoS measurements to detector choices under explicit application requirements.
We use detection time as an $\epsilon$-constraint~\cite{Haimes1971_EpsilonConstraint} while separately optimizing mistake rate or query accuracy, and instantiate this formulation to derive a detector-selection process for the evaluated operational space~(\Cref{sec:selection_process}).
\end{enumerate}

Together, these contributions have a broader implication for the evaluation of new failure detectors: because the best-performing detector varies with operating conditions and QoS objective, comparisons against a single or few baselines can give an incomplete picture of relative performance.
This is particularly relevant to learning-based failure detection, where improvements should be assessed against appropriate UFD alternatives at the same operating point rather than an arbitrarily chosen conventional detector.
Our benchmark and selection methodology provide a systematic way to identify such baselines.

\smallskip
To support reproducibility, we have publicly released our source code~\cite{ufd-eval}, including the data-collection and trace-replay components, detector implementations, and deployment scripts.
The collected trace dataset is also publicly available~\cite{dataset-2026-24465359}.

\section{\texorpdfstring{\(\Diamond\mathcal{P}\)}{Diamond P} and \texorpdfstring{\(\Diamond\mathcal{P}\)}{Diamond P}-Style Failure Detectors}\label{sec:sok_of_fds}

This section reviews the studied failure detectors.
Some are formal \(\Diamond\mathcal{P}\) detectors, while others provide closely related trust/suspect behavior or optimize QoS, overhead, or adaptation without proving \(\Diamond\mathcal{P}\) properties.
\Crefrange{sec:methods}{sec:results} evaluate them under a common benchmark.

\subsection{Monitoring Model and Detector Interfaces}\label{sec:model_and_detectors}
We adopt standard monitoring vocabulary.
A distributed system consists of nodes \(\Pi\), with monitoring nodes \(M \subseteq \Pi\) observing monitored nodes \(Q \subseteq \Pi\).
Nodes may fail by crashing and do not recover.
Detectors observe messages that may be delayed or lost.
When formal \(\Diamond \mathcal{P}\) guarantees are considered, the usual additional assumption is that communication delays eventually become bounded, although the bound does not need to be known~\cite{DBLP:journals/jacm/DworkLS88,DBLP:journals/jacm/ChandraT96}.

The studied detectors differ in both observation styles and output interfaces.
Push-style detectors~\cite{DBLP:journals/tc/ChenTA02a,DBLP:conf/dsn/BertierMS02,DBLP:conf/srds/HayashibaraDYK04,EDFD/Xiong07,WeibullFD/journal.pone.0173666,DBLP:conf/sac/SatzgerPTU07,DBLP:conf/ipps/TomsicSGAS15,2EFD/journal.pone.0191577,DBLP:journals/jpdc/SensARM24} rely on periodic heartbeat (HB) messages from monitored nodes.
Pull-style detectors~\cite{DBLP:conf/srds/DixitC10,DBLP:journals/sensors/DongRZL14,S-AFD/Liu15} actively send probes and wait for acknowledgments.
Most detectors output a binary trust/suspect value, either for a single monitored node or for a set of nodes.
Accrual detectors~\cite{DBLP:conf/srds/HayashibaraDYK04,EDFD/Xiong07,WeibullFD/journal.pone.0173666,DBLP:conf/sac/SatzgerPTU07} instead output a scalar suspicion level, leaving applications to choose a threshold for a binary decision.

\subsection{Chronological Evolution}\label{sec:chronology}
\begin{figure}[t]
  \centering
  \begin{tikzpicture}
    \node[anchor=south west, inner sep=0] (img)
    {\includegraphics[width=\linewidth]{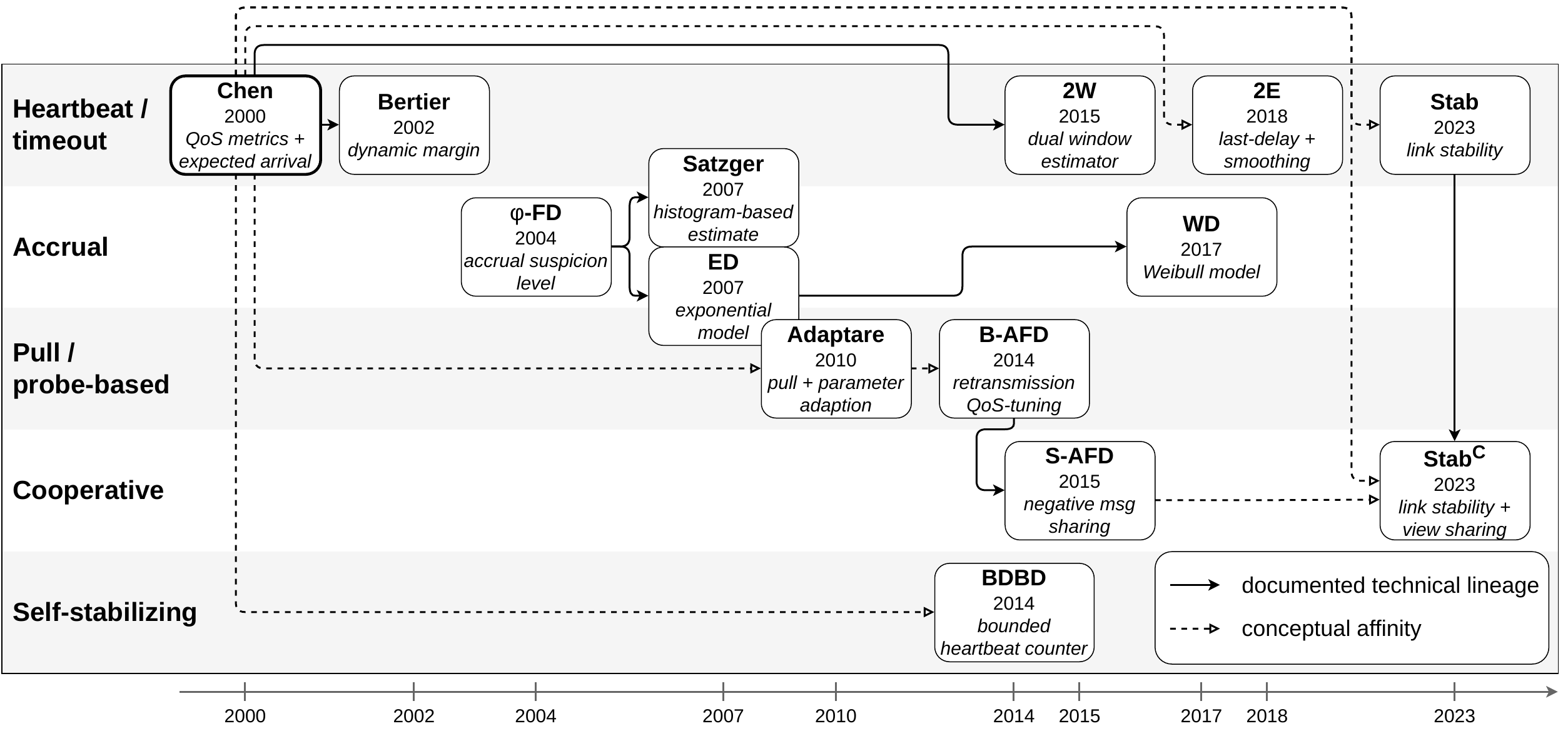}};

    \begin{scope}[x={(img.south east)}, y={(img.north west)}]
      \node[
        anchor=center,
        align=center,
        inner sep=0pt,
        outer sep=0pt,
        font=\fontsize{4}{5}\selectfont
      ] at (0.180,0.838)
        {\mbox{\cite{DBLP:journals/tc/ChenTA02a}}};
      \node[
        anchor=center,
        align=center,
        inner sep=0pt,
        outer sep=0pt,
        font=\fontsize{4}{5}\selectfont
      ] at (0.286,0.824)
        {\mbox{\cite{DBLP:conf/dsn/BertierMS02}}};
      \node[
        anchor=center,
        align=center,
        inner sep=0pt,
        outer sep=0pt,
        font=\fontsize{4}{5}\selectfont
      ] at (0.716,0.838)
        {\mbox{\cite{DBLP:conf/ipps/TomsicSGAS15}}};
      \node[
        anchor=center,
        align=center,
        inner sep=0pt,
        outer sep=0pt,
        font=\fontsize{4}{5}\selectfont
      ] at (0.834,0.838)
        {\mbox{\cite{2EFD/journal.pone.0191577}}};
      \node[
        anchor=center,
        align=center,
        inner sep=0pt,
        outer sep=0pt,
        font=\fontsize{4}{5}\selectfont
      ] at (0.953,0.824)
        {\mbox{\cite{DBLP:journals/jpdc/SensARM24}}};
        
      \node[
        anchor=center,
        align=center,
        inner sep=0pt,
        outer sep=0pt,
        font=\fontsize{4}{5}\selectfont
      ] at (0.368,0.672)
        {\mbox{\cite{DBLP:conf/srds/HayashibaraDYK04}}};
      \node[
        anchor=center,
        align=center,
        inner sep=0pt,
        outer sep=0pt,
        font=\fontsize{4}{5}\selectfont
      ] at (0.487,0.739)
        {\mbox{\cite{DBLP:conf/sac/SatzgerPTU07}}};
      \node[
        anchor=center,
        align=center,
        inner sep=0pt,
        outer sep=0pt,
        font=\fontsize{4}{5}\selectfont
      ] at (0.487,0.605)
        {\mbox{\cite{EDFD/Xiong07}}};
      \node[
        anchor=center,
        align=center,
        inner sep=0pt,
        outer sep=0pt,
        font=\fontsize{4}{5}\selectfont
      ] at (0.792,0.657)
        {\mbox{\cite{WeibullFD/journal.pone.0173666}}};
        
      \node[
        anchor=center,
        align=center,
        inner sep=0pt,
        outer sep=0pt,
        font=\fontsize{4}{5}\selectfont
      ] at (0.559,0.507)
        {\mbox{\cite{DBLP:conf/srds/DixitC10}}};
      \node[
        anchor=center,
        align=center,
        inner sep=0pt,
        outer sep=0pt,
        font=\fontsize{4}{5}\selectfont
      ] at (0.672,0.507)
        {\mbox{\cite{DBLP:journals/sensors/DongRZL14}}};

      \node[
        anchor=center,
        align=center,
        inner sep=0pt,
        outer sep=0pt,
        font=\fontsize{4}{5}\selectfont
      ] at (0.715,0.340)
        {\mbox{\cite{S-AFD/Liu15}}};
      \node[
        anchor=center,
        align=center,
        inner sep=0pt,
        outer sep=0pt,
        font=\fontsize{4}{5}\selectfont
      ] at (0.953,0.337)
        {\mbox{\cite{DBLP:journals/jpdc/SensARM24}}};

      \node[
        anchor=center,
        align=center,
        inner sep=0pt,
        outer sep=0pt,
        font=\fontsize{4}{5}\selectfont
      ] at (0.668,0.174)
        {\mbox{\cite{DBLP:conf/netys/BlanchardDBD14}}};
    \end{scope}
  \end{tikzpicture}
  
  \caption{Timeline and main relationships among the studied \(\Diamond\mathcal{P}\) and \(\Diamond\mathcal{P}\)-style detectors. Solid arrows denote documented technical lineage, while dashed arrows denote conceptual affinity. B-AFD is systematized but not implemented; we evaluate instead its extension S-AFD (\Cref{sec:chronology}).}
  \label{fig:fd_evolution_diagram}
\end{figure}

\Cref{fig:fd_evolution_diagram} presents the main relationships among the studied failure detectors.
For each detector, the timeline reports the earliest publication year we found, which may precede the peer-reviewed version.
A solid arrow denotes documented technical lineage, \ie, the target detector explicitly builds on or improves a prior detector.
A dashed arrow denotes conceptual affinity, \ie, the detectors use related techniques, but the articles do not state a direct lineage.
\Cref{tab:systematization_summary} complements the timeline by summarizing each detector's design dimension, output interface, and main refinement.

\subparagraph*{Heartbeat-Based Detection and Timeout Estimation.}
Chen~\etal~\cite{DBLP:journals/tc/ChenTA02a} connect \(\Diamond\mathcal{P}\)-style heartbeat detection with explicit QoS reasoning.
Their detector estimates the next heartbeat arrival from recent history, adds a constant safety margin, and suspects the monitored node if the heartbeat misses this freshness point.
Later heartbeat-based detectors refine this design by changing the arrival estimator, the safety margin, or the speed/accuracy trade-off.

Bertier~\cite{DBLP:conf/dsn/BertierMS02} keeps the expected-arrival estimator but replaces the constant safety margin with a Jacobson-style dynamic margin, allowing the timeout to react to recent estimation errors.
2W~\cite{DBLP:conf/ipps/TomsicSGAS15} maintains both a short-term and  long-term history window, using the short window to react to abrupt delay changes and the long window to retain stable-period information.
2E~\cite{2EFD/journal.pone.0191577} focuses on reducing estimator cost by using the last observed delay as the next-delay estimate and applying single exponential smoothing to the safety margin.
Stab~\cite{DBLP:journals/jpdc/SensARM24} calibrates the safety margin from observed link stability in WAN settings.
Its cooperative extension, Stab\(^C\), is discussed below.

\subparagraph*{Accrual Suspicion Level.}
\(\phi\)-FD~\cite{DBLP:conf/srds/HayashibaraDYK04} shifts from binary timeout decisions to an accrual interface.
Instead of returning trust or suspect, it outputs a continuous suspicion level derived from heartbeat inter-arrival behavior.
Applications then choose thresholds according to their own speed and accuracy requirements.
This separates monitoring from interpretation and allows one detector instance to support multiple application policies.

Satzger~\cite{DBLP:conf/sac/SatzgerPTU07}, ED~\cite{EDFD/Xiong07}, and WD~\cite{WeibullFD/journal.pone.0173666} preserve the accrual interface but change the statistical model used to compute suspicion.
Satzger estimates inter-arrival behavior with a histogram-based CDF.
ED uses an exponential inter-arrival model, while WD uses a Weibull model intended to better match unstable cloud-computing networks.
In this line, the main evolution is the mapping from delayed heartbeats to suspicion levels rather than the detector interface.

\subparagraph*{Probe-Based Detection.}
The pull, or probing, strategy predates Adaptare~\cite{DBLP:conf/srds/DixitC10}.
Bertier~\etal~\cite{DBLP:conf/dsn/BertierMS02} describe pinging as a classical alternative in which a monitor periodically sends ``are you alive?'' probes and waits for ``I am alive'' replies.
Among the detectors studied here, however, pull-style detection first becomes a central design choice in Adaptare.

Adaptare moves QoS-oriented detection from passive heartbeat monitoring to active probing.
It statistically characterizes the current network environment and adapts the interrogation period and timeout toward a configured QoS target under changing stochastic behavior.
B-AFD~\cite{DBLP:journals/sensors/DongRZL14} also uses pull-style detection, but adds a retransmission-based baseline model for lossy P2P networks.
Rather than treating one missed probe as sufficient evidence, it models repeated probes within a detection period so that QoS parameters can be tuned under message loss.
Since S-AFD~\cite{S-AFD/Liu15} extends B-AFD and reports better performance, our empirical evaluation focuses on S-AFD, see~\cite[Sec. IV]{S-AFD/Liu15}.

\subparagraph*{Cooperative Detection.}
S-AFD~\cite{S-AFD/Liu15} extends the pull-style line with cooperation to reduce system-level detection overhead.
Monitors still use B-AFD-style probing parameters, but selected host nodes share negative messages with other interested monitors when repeated failed probes indicate a crash.
Stab\(^C\)~\cite{DBLP:journals/jpdc/SensARM24} provides a heartbeat-based form of cooperation.
The nodes piggyback suspected-node sets and link-stability estimates, allowing monitors to compare their local observations with views from nodes that may have more stable links.
Cooperation therefore plays two roles in the studied detectors: reducing redundant probing in S-AFD and improving suspicion quality in Stab\(^C\).

\subparagraph*{Self-Stabilizing Failure Detection.}
Blanchard~\etal~\cite{DBLP:conf/netys/BlanchardDBD14} briefly describe the BDBD failure detector in the context of practically self-stabilizing Paxos.
Its motivation differs from the QoS-oriented detectors above.
Since transient faults may corrupt timestamp-like state, BDBD avoids timestamp comparisons and uses bounded heartbeat counters instead.
Each processor repeatedly broadcasts heartbeats, and each receiver keeps a bounded counter per processor, resets the sender's counter upon heartbeat receipt, and increments the other counters up to a fixed maximum \(W\).
A processor is reported as failed when its counter reaches \(W\).
BDBD is therefore a bounded-state heartbeat variant intended for a self-stabilizing setting, rather than a timeout-estimation algorithm.

\section{Evaluation Methodology} \label{sec:methods}

The studied detectors differ in interface, assumptions, and formal status.
We therefore compare them through their observable suspicion histories.
For detectors outputting binary decisions, we record trust/suspect transitions directly.
For accrual detectors, we apply explicit thresholds to convert suspicion levels into binary outputs.
These thresholds are treated as detector configurations.
This gives a common comparison interface across different designs.

\subsection{Trace-Based and Live Deployment Evaluation} \label{sec:trace_based_and_rt_evaluation}
Our framework uses two complementary evaluation modes: \emph{offline trace-based} and \emph{online live deployment} evaluation.
In offline evaluation, a data-collection process records network traces, and a simulator replays those traces to offline detector implementations.
This mode gives repeatable inputs and allows multiple detectors and configurations to be compared under the same recorded communication behavior.

In online evaluation, detectors are deployed directly on PlanetLab.
We use this mode for Adaptare, S-AFD, and Stab\(^C\), whose behavior depends on active probing or cooperation and therefore cannot be faithfully reconstructed from passive heartbeat traces alone.
All other detectors are evaluated offline through trace replay.

%
We use PlanetLab~\cite{PlanetLab/10.1145/956993.956995} to collect the network traces.
Each execution uses four nodes: one monitoring node and three monitored nodes.
The monitored nodes periodically send heartbeat messages to the monitoring node.
For each experiment configuration, data collection runs for 24 hours, producing a trace for offline replay.

%
Offline detectors are evaluated by replaying the collected traces through a simulator.
Replay preserves the recorded message-arrival behavior and allows the same trace corpus to be reused across detectors and parameter configurations.
For detectors that output binary trust/suspect values, the simulator records the resulting suspicion history directly.
For accrual detectors, we first apply an explicit threshold to convert the scalar suspicion level into a binary decision.
These thresholds are treated as detector configuration parameters.

%
Online detectors run on the same PlanetLab setup.
To make the duration comparable with trace collection, we match the total runtime, \ie, online detectors are executed sequentially for each configuration, with 12 configurations run for approximately two hours each, yielding about 24 hours of live observations in total.
This avoids overloading the PlanetLab nodes while still giving each online configuration a nontrivial observation window.

During online runs, we also collect the passive heartbeat stream used by the offline benchmark.
Conversely, during trace collection, we run the online detectors under the same deployment setup.
Thus, online and offline experiments use the same platform and comparable background host-load conditions, while online execution additionally includes the runtime effects of active probing, UDP reception, timer updates, and cooperative protocol processing.
These effects cannot be separated from algorithmic effects in the live measurements.

All components are implemented in Python~3.12, and all communication uses UDP via the standard \texttt{asyncio} and \texttt{socket} libraries.
Our detector implementations follow their original publications with specific design decisions, such as window size, described in \Cref{app:detector_implementation_details}.

\subsection{Crash Failure Simulation} \label{sec:crash_failure_simulation}
We evaluate responsiveness through repeated crash-stop measurement rounds.
In each round, a monitored node crashes by stopping all observable detector communication. 
The detector then records the elapsed time from crash onset to the first subsequent suspicion. 
After the round completes, the detector is reset and warmed up before the next round.
Thus, each round models a crash-stop failure, while repetition across rounds is used only to obtain multiple detection time measurements.

Failures are injected probabilistically.
We use a sampling probability chosen to produce an average inter-crash interval of about 30 seconds, with each injected crash lasting 30 seconds.
After each injected crash, detectors enter a reset and warmup phase before the next crash-injection round.
Metrics are not recorded during reset or warmup periods.

In offline experiments, failures are injected into replayed traces.
When a node is marked as crashed, all subsequent messages from that node are suppressed for the duration of the crash interval, so the replay reflects what the monitor would observe if the node stopped communicating.
Crashes are injected only after an initial warmup period, ensuring that detectors have enough prior observations to establish their normal operating state.

In live deployments, failures are injected by the running detector processes.
Each node independently samples whether to enter a crash interval, with the probability scaled by the number of nodes so that the system-wide crash frequency remains approximately one crash every 30 seconds.
During the crash interval, the node stops participating normally.
After the interval, the experiment enters reset and warmup before metric collection resumes.

The main distinction is therefore where the failure is applied.
Offline injection modifies an existing event trace by hiding communication from the crashed node, which gives repeatable and controlled experiments.
Crash-injection during live deployment affects live protocol execution, so nodes must coordinate crash-start and recovery events, and metric collection is paused during a warmup period after recovery to avoid measuring transient reset behavior.

\section{Evaluation Strategy}\label{sec:evaluation}
We decompose RQ1, which asks how the studied detectors perform under different operational conditions, into the three subquestions below. RQ2 concerns detector selection under specified constraints and is addressed in \Cref{sec:selection_process}.

\begin{description}
\item[RQ1.1]
How does the communication rate affect the latency/accuracy trade-off of \(\Diamond \mathcal{P}\) and \(\Diamond \mathcal{P}\)-style detectors, with and without frequent crash failures?

\item[RQ1.2]  
How does the host-side processing load affect detector mistake rates and query accuracy, and how does this interaction change under frequent crash failures?

\item[RQ1.3]
How do detectors that require active probing or cooperation behave in live deployment compared with offline detectors under comparable workloads and failure scenarios?

\end{description}

\subsection{Evaluation Criteria}\label{sec:evaluation_criteria}
We evaluate detectors using QoS metrics adapted from Chen~\etal~\cite{DBLP:journals/tc/ChenTA02a} and operationalized for our finite executions, repeated crash-injection rounds, and sampled detector histories.
\emph{Detection time} \(T_D\) is the elapsed time from an injected node crash until the detector first suspects the node.
\emph{Mistake rate} \(\lambda_M\) is the rate of false-suspicion transitions, \ie, transitions from trusting a correct node to incorrectly suspecting it.
\emph{Query accuracy probability} \(Q_A\) is the fraction of sampled times for which the detector output agrees with the known state of the monitored node.

\(T_D\) captures responsiveness, whereas \(\lambda_M\) and \(Q_A\) capture complementary aspects of accuracy.
In an environment with injected crash failures, \(T_D\) is measured from crash onset to the first suspicion.
When there is no crash failure to measure from, we estimate the worst-case \(T_D\) from the time when the last heartbeat was
sent until the detector's timeout (or threshold crossing for accrual detectors) expires.

We compute \(\lambda_M\) as the number of false-suspicion transitions divided by the effective metric-collection runtime.
We compute \(Q_A\) as the fraction of correct detector states among all query samples.
In live deployment, query accuracy is sampled once per second.
The metrics exclude the reset and warmup periods.
For readability, figures and tables report \(Q_E = 1 - Q_A\), so lower values indicate better performance for \(T_D\), \(\lambda_M\), and \(1 - Q_A\).

For a fixed scenario and error metric, we call a detector configuration optimal if no other evaluated configuration achieves both no larger \(T_D\) and no larger error, with a strict improvement in at least one of the two quantities.
The resulting set of optimal configurations defines the Pareto frontier, which we compute separately for \((T_D,\lambda_M)\) and \((T_D,1 - Q_A)\).

Offline replay can produce multiple values for the same metric across monitored nodes and independent crash-stop rounds.
We report the average over values obtained under the same detector, message rate, host-load setting, crash environment, and detector configuration.

\subsection{Evaluation Environment}\label{sec:evaluation_environment}
All experiments use four PlanetLab nodes: two in Sweden, one in Poland, and one in Belgium.
Each non-cooperative execution uses one monitoring node and three monitored nodes.
For cooperative detectors, all nodes perform both roles.
Trace collection and live deployment follow \Cref{sec:methods}.

We vary three experimental factors: communication rate, host load, and crash regime.
The communication rate is either \qty{100}{\Hz} or \qty{10}{\Hz}, corresponding to inter-message intervals of \qty{10}{\ms} and \qty{100}{\ms}.
For push-based detectors, this interval is the heartbeat interval, and for pull-based detectors, it is the probing interval.

We use three host-load settings: \emph{minimal}, \emph{moderate}, and \emph{noticeable}.
Under minimal load, nodes exchange only detector-related messages.
Under moderate and noticeable load, they also compute and attach host-load indicators, such as CPU and memory usage, adding approximately \qty{1}{\ms} and \qty{20}{\ms} of computation, respectively.

Crash behavior follows \Cref{sec:crash_failure_simulation}.
Offline detectors are evaluated under no-crash (NC) and crash-injected (CI) executions.
In particular, CI replays the same trace as NC with crash injection enabled.
NC isolates false suspicions, while CI measures behavior under repeated independent crash-stop rounds.
Online detectors are evaluated only under CI because live deployments are non-replayable and time-limited.
We allocate the online budget to CI to preserve per-configuration observation time while capturing both detection time and false suspicions during live probing or cooperation.

Detector configurations are selected to expose the trade-off between detection time and accuracy.
For timeout-based detectors, we vary parameters affecting timeout or detection latency; for accrual detectors, we vary the suspicion threshold converting scalar suspicion levels into binary decisions.
Online detectors use compact configuration sets so that each configuration can run for about two hours.
Detailed configurations are provided in \Cref{tab:detector_configs}.

Offline detectors are evaluated over the collected traces, while online configurations are executed sequentially to avoid overloading the nodes.
Two S-AFD results are omitted because extreme parameter values yielded no valid detection-time measurements.

\subsection{Experiment Plan}\label{sec:experiment_plan}
We organize the evaluation into three experiments using the QoS metrics from \Cref{sec:evaluation_criteria}.

\begin{description}
\item[Experiment 1.] 
Experiment 1 addresses RQ1.1. We evaluate the offline heartbeat-based and accrual detectors at \qty{100}{\Hz} and \qty{10}{\Hz} under NC and CI with moderate host load.
For each configuration, we measure average \(\lambda_M\) and \(1-Q_A\) as functions of \(T_D\), then compare the best attainable accuracy across detection-time thresholds.

\item[Experiment 2.]
Experiment 2 addresses RQ1.2.
We fix the heartbeat rate at \qty{100}{\Hz} and evaluate the same offline detectors under minimal, moderate, and noticeable host load in NC and CI, comparing \(\lambda_M\) and \(1-Q_A\) across detection-time thresholds.

\item[Experiment 3.]
Experiment 3 addresses RQ1.3.
We compare detectors requiring active probing or cooperation with Pareto-optimal offline configurations under moderate host load, CI, and \qty{100}{\Hz} and \qty{10}{\Hz} communication rates.
We use the same QoS metrics and detection-time thresholds as Experiments~1 and 2.

\end{description}

\section{Experimental Results}\label{sec:results}
We report the results of the three experiments from \Cref{sec:experiment_plan}.
We conclude each experiment with a boxed observation that states the corresponding answer to RQ1.

\subsection{Experiment 1: Impact of Heartbeat Rate}\label{sec:hb_rate_results}
\begin{figure}[ht]
    \centering
    \includegraphics[width=\linewidth]{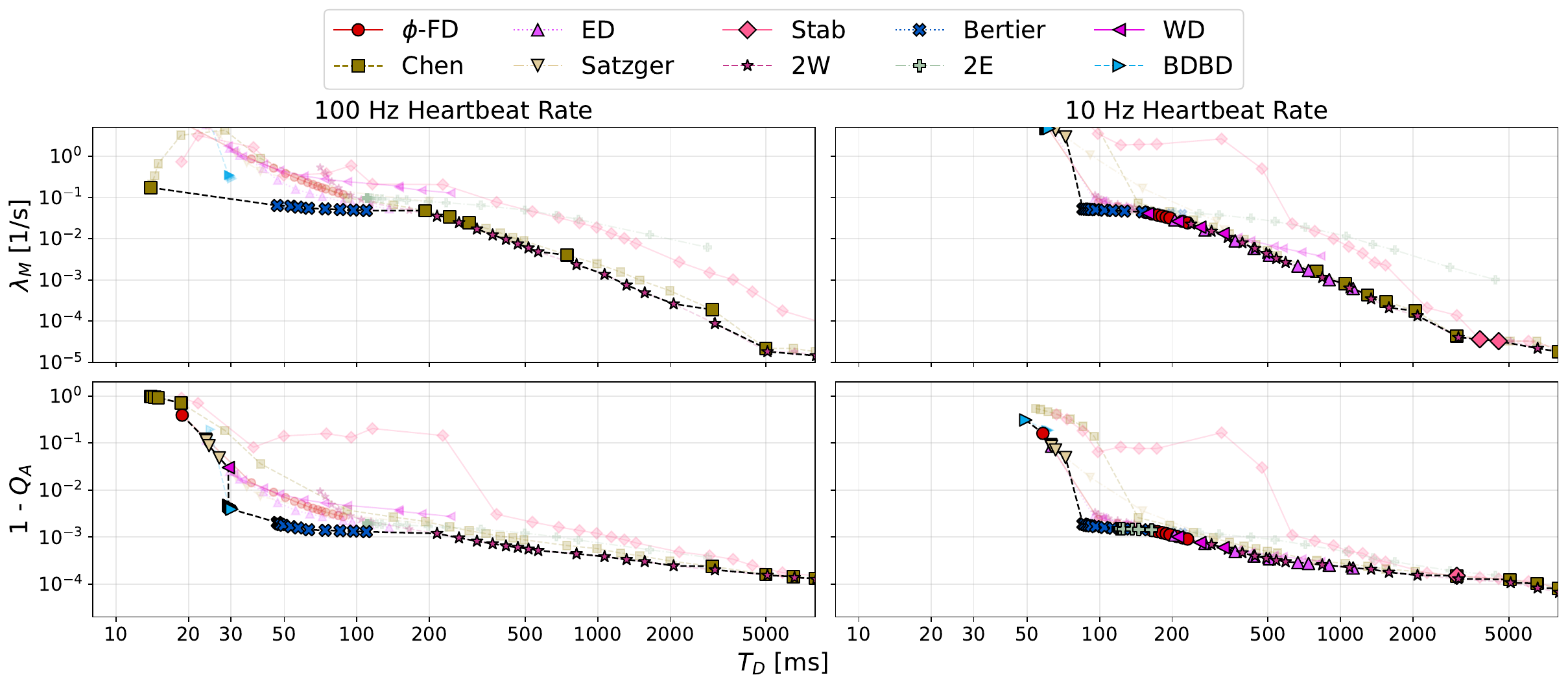}
    \caption{QoS metrics (\(\lambda_M\) and \(1 - Q_A\)) for offline detectors at \qty{100}{\Hz} and \qty{10}{\Hz} under CI environment. All axes use log scale.}
    \label{fig:hb_100hz_vs_10hz_CI}
\end{figure}

To answer RQ1.1, we compare offline detectors at \qty{100}{\Hz} and \qty{10}{\Hz} under no-crash (NC) and crash-injected (CI) environments.
\Cref{fig:hb_100hz_vs_10hz_CI} shows \(\lambda_M\) and \(1-Q_A\) as functions of \(T_D\) under CI environment.
Full plots for both CI and NC environments are provided in \Cref{app:detailed_results}, since they are visually similar.
We focus on configurations on the Pareto-frontier, since they capture the best observed trade-offs between detection speed and accuracy.

The expected latency/accuracy trade-off appears in both environments, with larger detection times generally corresponding to lower mistake rates and lower query error probabilities.
The heartbeat rate primarily affects the short detection times.
At \qty{100}{\Hz}, the earliest observed detections occur between \qty{10}{\ms} and \qty{20}{\ms}, whereas at \qty{10}{\Hz} they occur around \qty{50}{\ms}.
This follows from the shorter inter-heartbeat interval, which allows monitors to use shorter waiting periods before suspecting a node.

\Cref{tab:hb_best_results} lists the best \(\lambda_M\) and \(1-Q_A\) values under increasingly relaxed detection-time thresholds \(T_D \le 50, 100, 200, 500,\) and \(\qty{2000}{\ms}\).
Since NC and CI replay the same trace, a given configuration yields similar accuracy in both.
The main difference lies in short detection times.
NC estimates the worst-case \(T_D\), assuming a crash right after a heartbeat is sent, whereas an injected crash usually falls between two heartbeats and is detected sooner.
CI values therefore shift toward shorter \(T_D\), giving lower best-case \(\lambda_M\) and \(1-Q_A\) under strict thresholds, and the two converge once \(T_D\) is large relative to the heartbeat interval.
Because of this, NC and CI values are not directly comparable.

\begin{table}[ht]
\caption{Lowest \(\lambda_M\) and \(1 - Q_A\) achieved under each heartbeat rate, crash environment, and detection-time threshold.}
\label{tab:hb_best_results}
\centering
\begin{tblr}{
  colspec = {l c c c c c c c},
  colsep = 0.5em,
  rowsep = 0.15em,
  row{odd} = {bg=gray!10},
  row{1-2} = {font=\bfseries, halign=center, bg=white},
}
\toprule
\SetCell[r=2]{c} HB & \SetCell[r=2]{c} Env. & \SetCell[r=2]{c} Metric & \SetCell[c=5]{c} Detection Time Threshold & & & & \\
\cline{4-8}
& & & $T_D \le 50$ & $T_D \le 100$ & $T_D \le 200$ & $T_D \le 500$ & $T_D \le 2000$ \\
\midrule
\qty{100}{\Hz} & CI & $\lambda_M$ & 0.064 & 0.049 & 0.048 & 0.0074  & \num{4.9e-4} \\
\qty{10}{\Hz} & CI & $\lambda_M$ & 7.51 & 0.050 & 0.032 & 0.0044  & \num{2.1e-4} \\
\qty{100}{\Hz} & CI & $1 - Q_A$ & 0.0018 & 0.0013 & 0.0013 & \num{6.0e-4}  & \num{3.0e-4}  \\
\qty{10}{\Hz} & CI & $1 - Q_A$ & 0.31 & 0.0017  & 0.0011 & \num{3.5e-4}  & \num{1.8e-4}  \\
\midrule
\qty{100}{\Hz} & NC & $\lambda_M$ & 0.173 & 0.072 & 0.048 & 0.0074  & \num{4.9e-4} \\
\qty{10}{\Hz} & NC & $\lambda_M$ & N.A. & 7.27 & 0.047 & 0.0058  & \num{2.1e-4} \\
\qty{100}{\Hz} & NC & $1 - Q_A$ & 0.0039 & 0.0021  & 0.0013 & \num{6.0e-4}  & \num{3.0e-4}  \\
\qty{10}{\Hz} & NC & $1 - Q_A$ & N.A. & 0.16  & 0.0016 & \num{4.0e-4}  & \num{1.8e-4}  \\

\bottomrule
\end{tblr}
\end{table}

For strict detection thresholds, \qty{100}{\Hz} improves accuracy substantially.
In NC at \(T_D \le \qty{100}{\ms}\), \qty{10}{\Hz} gives \(\lambda_M = 7.27\) and \(1-Q_A = 0.16\), while \qty{100}{\Hz} gives \(\lambda_M = 0.072\) and \(1-Q_A = 0.0021\), corresponding to reductions of 99\% and 98.7\%.
This advantage diminishes once \(T_D \ge \qty{200}{\ms}\).
At \(T_D \le \qty{500}{\ms}\) in NC, \qty{10}{\Hz} reaches \(\lambda_M = 0.0058\) and \(1-Q_A = \num{4.0e-4}\), compared with \(\lambda_M = 0.0074\) and \(1-Q_A = \num{6.0e-4}\) at \qty{100}{\Hz}.
Thus, when applications can tolerate longer detection times, a lower heartbeat rate can be comparable to or better.

\begin{observation}[RQ1.1 -- communication-rate effect] \label{obs:communication-rate-effect}
In our traces, \qty{100}{\Hz} significantly improves accuracy primarily under strict detection-time budgets.
In NC at  \(T_D \leq \qty{100}{\ms}\), increasing the heartbeat rate from \qty{10}{\Hz} to \qty{100}{\Hz} reduces \(\lambda_M\) by 99.0\% and \(1 - Q_A\) by 98.7\%.
Once the detection-time budget is relaxed, this advantage disappears.
In NC at  \(T_D \leq \qty{500}{\ms}\), \qty{10}{\Hz} reduces \(\lambda_M\) by approximately 22\% and \(1-Q_A\) by 33\% relative to \qty{100}{\Hz}.
Thus, a higher message rate mainly improves accuracy in the short-latency operating environment rather than uniformly improving attainable accuracy.
\end{observation}

\subsection{Experiment 2: Impact of Host Load}\label{sec:host_load_impact}
\begin{figure}
    \centering
    \includegraphics[width=\linewidth]{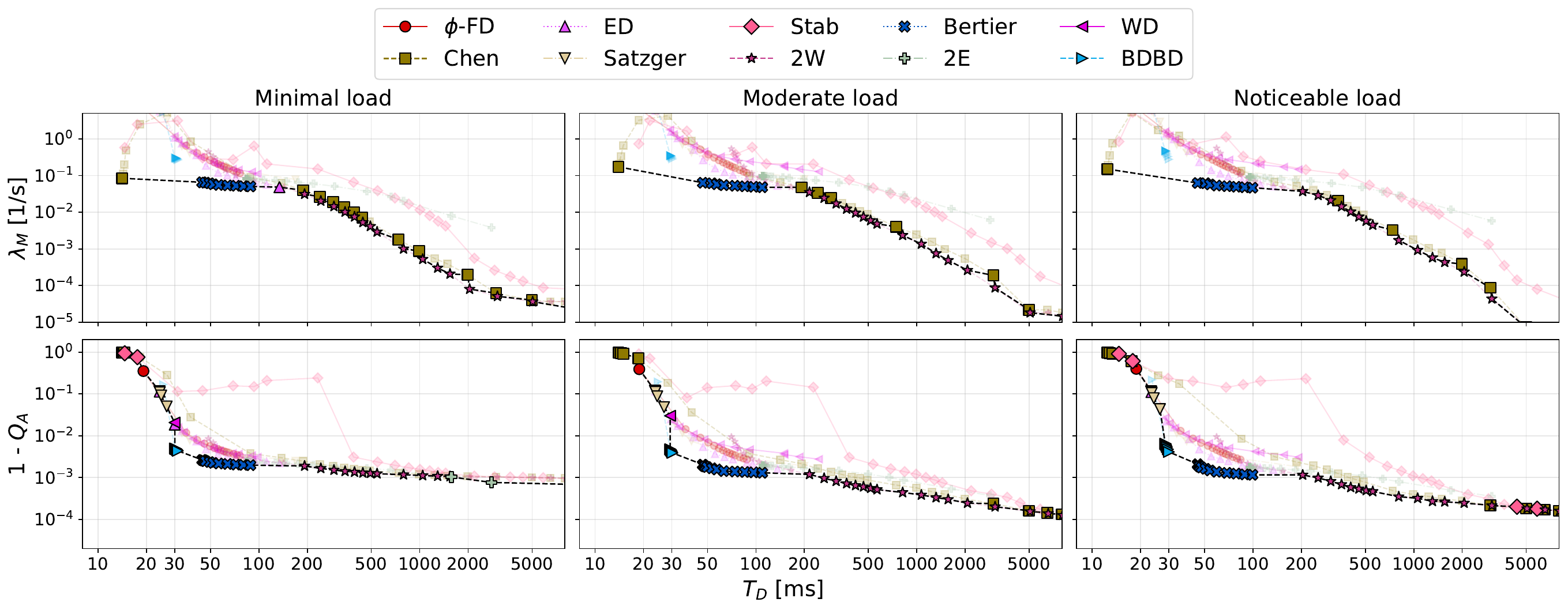}
    \caption{QoS metrics (\(\lambda_M\) and \(1 - Q_A\)) for offline detectors under minimal, moderate, and noticeable host load under CI environment. All axes use log scale.}
    \label{fig:host_load_comparisons_CI}
\end{figure}

To answer RQ1.2, we fix the heartbeat rate at \qty{100}{\Hz} and compare minimal, moderate, and noticeable host-load settings.
\Cref{fig:host_load_comparisons_CI} reports \(\lambda_M\) and \(1-Q_A\) under CI.
The full results for both CI and NC are found in \Cref{app:detailed_results}.
We use \qty{100}{\Hz} because shorter heartbeat intervals make host-side processing effects more visible.

\begin{table}[ht]
\centering
\caption{Summary statistics of heartbeat inter-arrival times in milliseconds under different host loads and \qty{100}{\Hz} heartbeat rate.}
\label{tab:inter_arrival_summary}
\begin{tblr}{
  colspec={lrrrrrrr},
  colsep = 0.5em,
  rowsep = 0.15em,
  row{even} = {bg=gray!10},
  row{1} = {font=\bfseries, halign=center},
}
\toprule
Load & Mean & Median & Std. dev. & p99 & p99.9 & Min & Max \\
\midrule
minimal &
10.11 &
10.08 &
43.61 &
17.95 &
67.27 &
0.04 &
\num{219720}
\\
moderate &
10.14 &
10.08 &
14.43 &
20.95 &
77.69 &
0.05 &
\num{25423}
\\
noticeable &
10.13 &
10.07 &
10.99 &
20.54 &
71.46 &
0.04 &
\num{4267} 
\\
\bottomrule
\end{tblr}
\end{table}

Moderate and noticeable load behave similarly in both environments, whereas minimal load differs.
In particular, the minimal-load query-error curves flatten earlier, around \(1-Q_A = 10^{-3}\), while the other loads flatten around \(10^{-4}\).
This is not because heartbeats arrive more irregularly in general, since \Cref{tab:inter_arrival_summary} shows similar means and medians, and the minimal-load trace even has the lowest p99 and p99.9 values.
Its larger standard deviation instead stems from a single \qty{219.7}{\second} gap, during which the node is falsely suspected, resulting in the earlier plateau in \(1 - Q_A\).
Since it causes only one false suspicion, it barely affects \(\lambda_M\).

\Cref{tab:load_best_results} reports the best \(\lambda_M\) and \(1-Q_A\) under NC and CI environments for the detection-time threshold \(T_D \le \qty{2000}{\ms}\).
We omit shorter thresholds because they follow the trends observed in earlier experiments: 
CI mainly attains lower values at short detection times, due to the worst-case \(T_D\) estimate used in NC (\Cref{sec:hb_rate_results}).

At \(T_D \le \qty{2000}{\ms}\), \(1-Q_A\) is about \num{0.001} under minimal load, compared with
\num{3.0e-4} and \num{2.6e-4} under moderate and noticeable load, consistent with the gap observed above.
In contrast, \(\lambda_M\) is up to 145\% higher under moderate and noticeable load.
For even larger \(T_D\), \Cref{fig:host_load_comparisons_CI} shows the minimal-load \(\lambda_M\) curve flattening above \(10^{-5}\), while the other settings converge around or below \(10^{-5}\).

\begin{table}[t]
\caption{Lowest mistake rates \(\lambda_M\) and query error probabilities \(1 - Q_A\) achieved under each host-load setting, crash environment, and detection time threshold \(T_D \le \qty{2000}{\ms}\).}
\label{tab:load_best_results}
\centering
\begin{tblr}{
  colspec = {l l c c},
  rowsep = 0.15em,
  row{even} = {bg=gray!10},
  row{1} = {font=\bfseries, halign=center, bg=white},
}
\toprule
Load & Env. & \(\bm{\lambda_M}\) & \(\bm{1-Q_A}\) \\
\midrule
minimal   & CI & $2.0 \cdot 10^{-4}$ & $0.0010$ \\
moderate  & CI & $4.9 \cdot 10^{-4}$ & $3.0 \cdot 10^{-4}$ \\
noticeable & CI & $3.9 \cdot 10^{-4}$ & $2.6 \cdot 10^{-4}$ \\
\midrule
minimal   & NC & $2.0 \cdot 10^{-4}$ & $0.0011$ \\
moderate  & NC & $4.9 \cdot 10^{-4}$ & $3.0 \cdot 10^{-4}$ \\
noticeable & NC & $3.9 \cdot 10^{-4}$ & $2.6 \cdot 10^{-4}$ \\
\bottomrule
\end{tblr}
\end{table}

\begin{observation}[RQ1.2 -- host-side load effect] \label{obs:host-load-effect}

At the configured host loads, we observe no consistent effect on attainable accuracy.
At \(T_D \le \qty{2000}{\ms}\), the moderate- and noticeable-load traces show lower query error (up to \(74\%\) in CI, \(76\%\) in NC), but this difference is attributable to a single \qty{219.7}{\second} gap in the minimal-load trace rather than to host load.
Mistake rates show the opposite ordering (up to \(145\%\) higher).
Rare long gaps thus dominate the host-load effect, and \(1-Q_A\) is far more sensitive to them than \(\lambda_M\).
\end{observation} 

\subsection{Experiment 3: Comparison of Online and Offline Detectors}\label{sec:online_vs_offline_detectors}
\begin{figure}[ht]
    \centering
    \includegraphics[width=\linewidth]{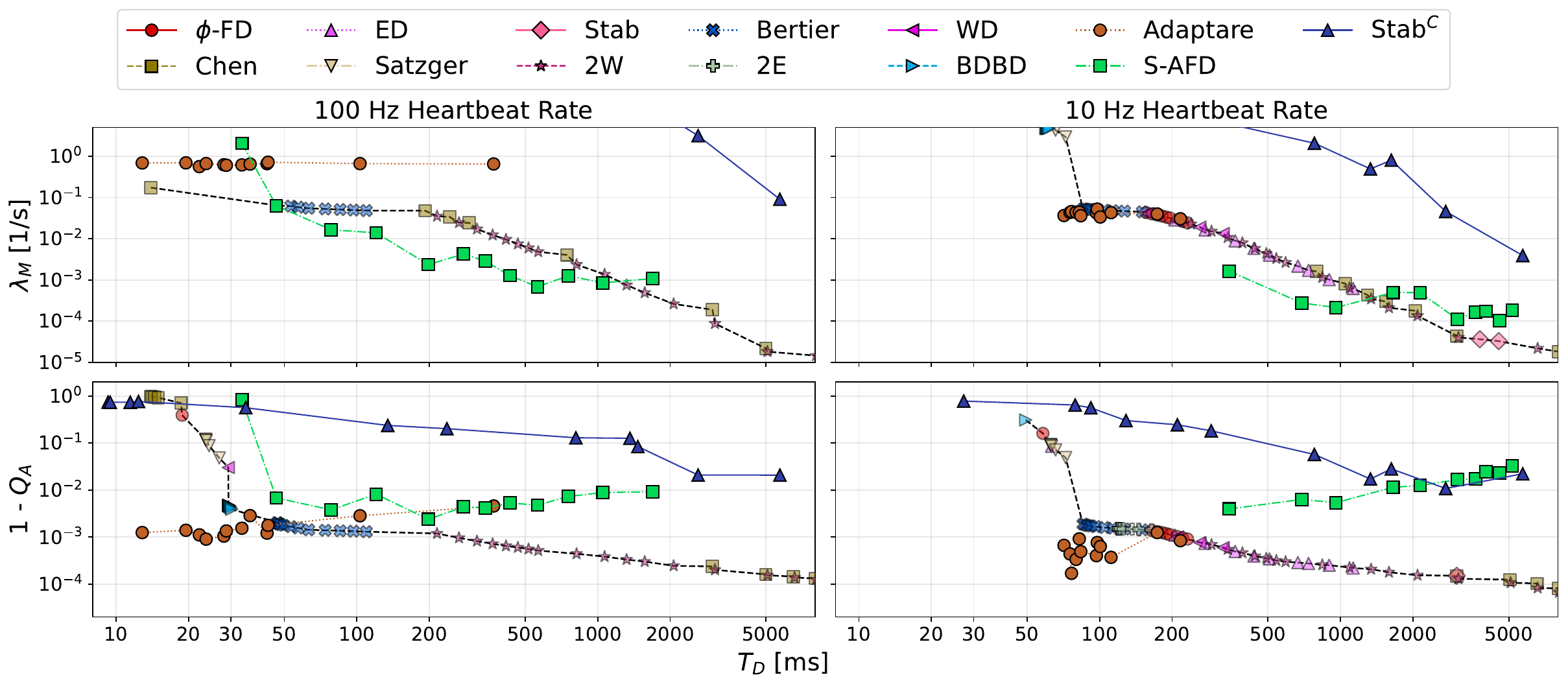}
    \caption{QoS metrics (\(\lambda_M\) and \(1 - Q_A\)) for online detectors and Pareto-optimal offline detectors at \qty{100}{\Hz} and \qty{10}{\Hz} under CI. All axes use log scale.}
    \label{fig:real_time_fd}
\end{figure}

To answer RQ1.3, we compare the online detectors (Adaptare, S-AFD, and Stab\(^C\)) with the Pareto-optimal offline detectors under moderate host load and CI, see \Cref{fig:real_time_fd}.

At \qty{100}{\Hz}, S-AFD performs well on mistake rate.
It attains lower observed \(\lambda_M\) than most offline detectors for \(T_D > \qty{50}{\ms}\) and degrades mainly for \(T_D > \qty{1000}{\ms}\).
At \(T_D \leq \qty{500}{\ms}\), S-AFD reduces \(\lambda_M\) by more than 80\% relative to the second-best detector.
However, it does not improve query error relative to the strongest offline detectors.

Adaptare shows the opposite trade-off at \qty{100}{\Hz}.
It provides fast detection and low query error, with approximately \(1-Q_A = 0.0013\) at \(T_D \approx \qty{15}{\ms}\), but its mistake rate remains high for most tested configurations, around \(\lambda_M \approx 0.6\).
Stab\(^C\) has mistake rates and query error probabilities that are often an order of magnitude higher than those of other detectors at similar detection times, showing that it does not perform well under our setting.

These differences may reflect both the active or cooperative algorithms themselves and the execution effects present in live deployment.
At \qty{100}{\Hz}, scheduling delays and network jitter are large relative to the message interval and can produce premature timeouts or frequent false suspicions.
S-AFD may be less affected because its parameters control retry spacing within a detection period rather than a sustained decision rate.
Namely, it sends a bounded number of probes, stops after the first acknowledgment, and suspects only after repeated failures within the detection period.

At \qty{10}{\Hz}, all three online detectors show lower mistake rates.
S-AFD remains strong in terms of \(\lambda_M\), reaching below \(10^{-3}\) for \(T_D > \qty{500}{\ms}\), but its earliest detection time increases to approximately \qty{340}{\ms}.
Its query error also worsens as detection time increases, likely because longer detection periods extend incorrect-state durations.
Stab\(^C\) slightly improves its mistake rate but remains worse than stronger alternatives.

Adaptare performs particularly well among the online detectors at \qty{10}{\Hz}.
It achieves \(1-Q_A\) between \(10^{-4}\) and \(10^{-3}\) for most configurations, with detection times around \qtyrange{70}{150}{\ms}.
Its mistake rates are also comparable to or better than the Pareto-optimal offline detectors at similar detection times.
Relative to the offline detectors at \(T_D \leq \qty{100}{\ms}\), Adaptare reduces  \(1 - Q_A\) by up to approximately 90\% while maintaining comparable mistake~rates.

\begin{observation}[RQ1.3 -- online detectors] \label{obs:online-detectors}
In the live CI experiments, S-AFD provides the strongest mistake-rate behavior at \qty{100}{\Hz}, reducing \(\lambda_M\) by more than 80\% at \(T_D \leq \qty{500}{\ms}\), whereas Adaptare performs strongly at \qty{10}{\Hz} around \(T_D \leq \qty{100}{\ms}\), reducing query error by up to approximately 90\% while maintaining comparable mistake rates.
No live detector is uniformly best across message rates and accuracy metrics.
Because these detectors are executed live rather than replayed on identical traces, the comparison characterizes their deployed behavior under comparable conditions, but does not isolate algorithmic effects from live scheduling and network jitter.
\end{observation} 

\section{From Benchmark Results to Detector Selection}\label{sec:selection_process}

RQ2 asks how empirical QoS trade-offs can be translated into suitable detector candidates under specified deployment and application constraints.
We formulate this as Rice's \emph{Algorithm Selection Problem}, which selects an algorithm from a portfolio based on problem-instance characteristics and observed algorithm performance~\cite{DBLP:journals/ac/Rice76,DBLP:journals/aim/Kotthoff14,DBLP:journals/ec/KerschkeHNT19}. Here, the portfolio is the studied failure detectors, and an instance is characterized by deployment conditions such as communication rate, host-load setting, and crash regime. We use this framework as a decision model for the evaluated deployment scenarios.

\subsection{Selection Method}
\label{sec:selection_method}

Failure-detector selection involves competing QoS objectives: shorter detection time can trade off against accuracy.
We therefore apply the \(\epsilon\)-constraint approach~\cite{Haimes1971_EpsilonConstraint}, treating the application's maximum acceptable detection time as a constraint and optimizing an accuracy metric among configurations satisfying it rather than combining detection time and accuracy in an arbitrary weighted score.

Let \(s = (r, l, c)\) denote an evaluated deployment scenario, characterized by communication rate \(r\), host load \(l\), and crash regime \(c\).
Let \(\mathcal{A}\) be the portfolio of benchmarked detector algorithms, and let \(\Theta_a(s)\) be the set of configurations of detector \(a\) evaluated in scenario $s$.
The available candidate set is therefore \(C_s = \{(a,\theta): a \in \mathcal{A}, \theta \in \Theta_a(s)\}\).
Let \(B\) be the maximum acceptable
detection time and \(m \in \{\lambda_M, 1-Q_A\}\) be the application's accuracy error metric to minimize.
We select \(X^*(s,B,m)=\left({\arg\min}_{x\in\mathcal{C}_{s}:T_D(x)\le B}\ m(x)\right)\).
Thus, \(X^*(s,B,m)\) is the detector--configuration pair that minimizes the chosen accuracy error metric \(m\) among all candidates whose detection time does not exceed \(B\).

Equivalently, the search may be restricted to the corresponding Pareto frontier because a dominated configuration cannot improve the constrained objective. 
Each \(x\) is a detector–configuration pair. Since RQ2 concerns detector choice rather than parameter tuning, \Cref{tab:best_fds_realtime_summary} reports the detector identity associated with the selected pair, while the benchmark retains the configuration required to reproduce the measurement.

\subsection{Applying the Method to Our Benchmark}
\label{sec:applying_selection_method}

We apply the selection method to the scenarios evaluated in \Cref{sec:results}.
For each scenario and accuracy metric, we
(i) compute the corresponding Pareto frontier,
(ii) retain points satisfying the detection-time threshold \(T_D\le B\), and
(iii) report the detector associated with the lowest remaining error.
We apply this procedure for
\(B\in\{50,100,200,500,\qty{2000}{\ms}\}\).
If no evaluated configuration satisfies a budget, the result is reported as N.A.
\Cref{tab:best_fds_realtime_summary} reports the resulting detector-level selections.
Notably, no evaluated case yielded multiple candidates, so we do not specify a tie-breaking convention.

\begin{table}[t]
\caption{Detector-level selection results by scenario, accuracy metric, and detection-time budget.
For each threshold, the table reports the detector with the lowest error among Pareto-optimal configurations satisfying the detection-time constraint.
NC scenarios use the worst-case \(T_D\) estimate (\Cref{sec:evaluation_criteria}).
}
\label{tab:best_fds_realtime_summary}
\centering
\small
\begin{tblr}{
  colspec = {l c c c c c c},
  colsep = 0.5em,
  rowsep = 0.15em,
  row{odd} = {bg=gray!10},
  row{1} = {font=\bfseries, halign=center, bg=white},
}
\toprule
\SetCell[r=2]{c} Scenario & \SetCell[r=2]{c} Metric & \SetCell[c=5]{c} Detection Time Threshold & & & & \\
\cline{3-7}
& & $T_D \le 50$ & $T_D \le 100$ & $T_D \le 200$ & $T_D \le 500$ & $T_D \le 2000$ \\
\midrule

100-min-CI & $\lambda_M$ & Bertier & Bertier & 2W & 2W & Chen \\
           & $1-Q_A$     & Bertier & Bertier & 2W & 2W & 2E \\

100-mod-CI & $\lambda_M$ & S-AFD & S-AFD & S-AFD & S-AFD & 2W \\
           & $1-Q_A$     & Adaptare & Adaptare & Adaptare & 2W & 2W \\

10-mod-CI  & $\lambda_M$ & BDBD & Adaptare & \(\phi\)-FD & S-AFD & 2W \\
           & $1-Q_A$     & BDBD & Adaptare & Adaptare & Adaptare & Adaptare \\

100-not-CI & $\lambda_M$ & Bertier & Bertier & Bertier & 2W & Chen \\
           & $1-Q_A$     & Bertier & Bertier & Bertier & 2W & 2W \\
100-min-NC & $\lambda_M$ & Chen & ED & Chen & Chen & Chen \\
           & $1-Q_A$     & BDBD & ED & Bertier & 2W & 2W \\
100-mod-NC & $\lambda_M$ & Chen & ED & Chen & 2W & 2W \\
           & $1-Q_A$     & BDBD & ED & Bertier & 2W & 2W \\
10-mod-NC  & $\lambda_M$ & N.A. & \(\phi\)-FD & \(\phi\)-FD & 2W & 2W \\
           & $1-Q_A$     & N.A. & \(\phi\)-FD & \(\phi\)-FD & 2W & 2W \\
100-not-NC & $\lambda_M$ & Chen & ED & Bertier & 2W & Chen \\
           & $1-Q_A$     & BDBD & ED & Bertier & 2W & 2W \\
\bottomrule
\end{tblr}
\end{table}

Using \Cref{tab:best_fds_realtime_summary}, the selection is direct once an application specifies its detection-time threshold and accuracy metric.
As Chen~\etal~\cite{DBLP:journals/tc/ChenTA02a} note, different applications care about different aspects of accuracy: \(\lambda_M\) matters for long-lived applications in which each mistake triggers a costly interrupt, such as group membership and cluster management, whereas \(Q_A\) matters for applications that query the detector at random times. 
Since neither metric alone fully characterizes accuracy, \Cref{tab:best_fds_realtime_summary} reports both.

Applying the selection rule yields two recurring patterns for our results.
First, at higher detection-time thresholds, 2W is the most frequently selected detector: in 12 of 16 scenario--metric combinations at \(T_D\le \qty{500}{\ms}\) and 10 of 16 at \(T_D\le \qty{2000}{\ms}\).
Second, under stricter detection-time budgets (\(B \le \qty{200}{\ms}\)), no detector is uniformly preferred: the selected candidate depends on the evaluated deployment conditions and, in several cases, on whether the application minimizes mistake rate or query error.

\subparagraph*{Answer to RQ2.}
We formulate failure-detector selection as a constraint-based algorithm-selection challenge~\cite{DBLP:journals/ac/Rice76}.
Given an operational scenario, a detection-time threshold, and an accuracy objective, we restrict the corresponding Pareto frontier to feasible configurations and select the detector with minimum error.
Applying this method to our benchmark yields the detector-level selections in \Cref{tab:best_fds_realtime_summary}.
In particular, 2W is selected most frequently at relaxed detection-time thresholds, whereas stricter thresholds yield scenario- and metric-dependent choices.
The same selection procedure can be rerun using benchmark measurements collected from other deployment environments.

\section{Conclusion}\label{sec:conclusion}
Three decades after Chandra and Toueg introduced unreliable failure detectors~\cite{DBLP:journals/jacm/ChandraT96}, many practical implementations have emerged, but their evaluations remain fragmented across environments, workloads, parameterizations, and comparison sets.
Until now, this fragmentation hindered both detector selection for a given deployment and the establishment of strong analytical baselines for newer approaches, including ML-based detectors.

We address this gap through a combined systematization, empirical benchmark, and detector-selection methodology.
We systematize 14 $\Diamond \mathcal{P}$ and $\Diamond \mathcal{P}$-style detectors and implement 13 of them under a common QoS evaluation framework, comparing detection time and accuracy.
We formalize detector selection as a constraint-based algorithm-selection challenge~\cite{DBLP:journals/ac/Rice76}, mapping measured QoS trade-offs to candidates under explicit deployment conditions, detection-time requirements, and accuracy objectives.
We demonstrate rapid deployment of this methodology on PlanetLab, combining same-input trace replay with live experiments for detectors requiring online execution for probing or cooperation.
Our results reinforce that detector selection should be conditioned on operational requirements rather than reduced to a universal ranking.
No detector is uniformly preferable across the evaluated detection-time and accuracy objectives: relaxed detection-time thresholds yield more consistent choices, whereas stricter thresholds require scenario- and metric-specific selection.
Our benchmarking framework and selection procedure can be rerun in other deployment environments, providing both a reusable basis for practical UFD selection and a broader baseline for evaluating future failure-detection approaches, including learning-based detectors.

\bibliography{references}

@misc{dataset-2026-24465359,
  author       = {Wang, Yenan},
  title        = {{PlanetLab} Network Traces 260609},
  month        = nov,
  year         = 2026,
  publisher    = {Zenodo},
  howpublished = {Zenodo},
  doi          = {10.5281/zenodo.24465359},
  url          = {https://doi.org/10.5281/zenodo.24465359},
}

@misc{ufd-eval,
  author       = {Wang, Yenan and others},
  title        = {{ufd-eval}: Evaluation of Unreliable Failure Detectors},
  year         = {2026},
  howpublished = {\url{https://bitbucket.org/eladschiller/ufd-eval/}},
  note         = {Bitbucket repository, commit \texttt{	
c039952}}
}

@article{DBLP:journals/jsac/XiongVYSPKL09,
  author       = {Naixue Xiong and
                  Athanasios V. Vasilakos and
                  Laurence Tianruo Yang and
                  Lingyang Song and
                  Yi Pan and
                  Rajgopal Kannan and
                  Yingshu Li},
  title        = {Comparative analysis of quality of service and memory usage for adaptive
                  failure detectors in healthcare systems},
  journal      = {{IEEE} J. Sel. Areas Commun.},
  volume       = {27},
  number       = {4},
  pages        = {495--509},
  year         = {2009},
  doi          = {10.1109/JSAC.2009.090512},
}

@inproceedings{DBLP:conf/dsn/DefagoUHK05,
  author       = {Xavier D{\'{e}}fago and
                  P{\'{e}}ter Urb{\'{a}}n and
                  Naohiro Hayashibara and
                  Takuya Katayama},
  title        = {Definition and Specification of Accrual Failure Detectors},
  booktitle    = {2005 International Conference on Dependable Systems and Networks {(DSN}
                  2005), 28 June - 1 July 2005, Yokohama, Japan, Proceedings},
  pages        = {206--215},
  publisher    = {{IEEE} Computer Society},
  year         = {2005},
  doi          = {10.1109/DSN.2005.37},
}

@article{DBLP:journals/ijpcc/FriedmanT09,
  author       = {Roy Friedman and
                  Galya Tcharny},
  title        = {Evaluating failure detection in mobile ad-hoc networks},
  journal      = {Int. J. Pervasive Comput. Commun.},
  volume       = {5},
  number       = {4},
  pages        = {476--496},
  year         = {2009},
  doi          = {10.1108/17427370911008857},
}

@article{DBLP:journals/ppl/KumarW19,
  author       = {Saptaparni Kumar and
                  Jennifer L. Welch},
  title        = {Implementing {$\diamondsuit$P} with Bounded Messages on a Network
                  of \emph{ADD} Channels},
  journal      = {Parallel Process. Lett.},
  volume       = {29},
  number       = {1},
  pages        = {1950002:1--1950002:12},
  year         = {2019},
  doi          = {10.1142/S0129626419500026},
}

@inproceedings{DBLP:conf/netys/RumreichS22,
  author       = {Laine E. Rumreich and
                  Paolo A. G. Sivilotti},
  title        = {An Eventually Perfect Failure Detector on {ADD} Channels Using Clustering},
  booktitle    = {Networked Systems - 10th International Conference, {NETYS} 2022, Virtual
                  Event, May 17-19, 2022, Proceedings},
  series       = {Lecture Notes in Computer Science},
  volume       = {13464},
  pages        = {149--166},
  publisher    = {Springer},
  year         = {2022},
  doi          = {10.1007/978-3-031-17436-0_11},
}

@inproceedings{DBLP:conf/ladc/SteinRDA23,
  author       = {Gabriela Stein and
                  Luiz Antonio Rodrigues and
                  Elias Proc{\'{o}}pio Duarte Jr. and
                  Luciana Arantes},
  title        = {{D}iamond-{P}-v{C}ube: An Eventually Perfect Hierarchical Failure Detector
                  for Asynchronous Distributed Systems},
  booktitle    = {12th Latin-American Symposium on Dependable and Secure Computing,
                  {LADC} 2023, La Paz, Bolivia, October 16-18, 2023},
  pages        = {40--49},
  publisher    = {{ACM}},
  year         = {2023},
  url          = {https://doi.org/10.1145/3615366.3615420},
  doi          = {10.1145/3615366.3615420},
}

@article{DBLP:journals/ppl/VargasRR20,
  author       = {Karla Vargas and
                  Sergio Rajsbaum and
                  Michel Raynal},
  title        = {An Eventually Perfect Failure Detector for Networks of Arbitrary Topology
                  Connected with {ADD} Channels Using Time-To-Live Values},
  journal      = {Parallel Process. Lett.},
  volume       = {30},
  number       = {2},
  pages        = {2050006:1--2050006:23},
  year         = {2020},
  doi          = {10.1142/S0129626420500061},
}

@article{DBLP:journals/ec/KerschkeHNT19,
  author       = {Pascal Kerschke and
                  Holger H. Hoos and
                  Frank Neumann and
                  Heike Trautmann},
  title        = {Automated Algorithm Selection: Survey and Perspectives},
  journal      = {Evol. Comput.},
  volume       = {27},
  number       = {1},
  pages        = {3--45},
  year         = {2019},
  doi          = {10.1162/EVCO_A_00242},
}

@inproceedings{DBLP:conf/usenix/GuerraouiMPXYZ22,
  author       = {Rachid Guerraoui and
                  Antoine Murat and
                  Javier Picorel and
                  Athanasios Xygkis and
                  Huabing Yan and
                  Pengfei Zuo},
  title        = {u{K}haron: {A} Membership Service for Microsecond Applications},
  booktitle    = {Proceedings of the 2022 {USENIX} Annual Technical Conference, {USENIX}
                  {ATC} 2022, Carlsbad, CA, USA, July 11-13, 2022},
  pages        = {101--120},
  publisher    = {{USENIX} Association},
  year         = {2022},
}

@inproceedings{DBLP:conf/usenix/SureshMGCL18,
  author       = {Lalith Suresh and
                  Dahlia Malkhi and
                  Parikshit Gopalan and
                  Ivan Porto Carreiro and
                  Zeeshan Lokhandwala},
  title        = {Stable and Consistent Membership at Scale with {R}apid},
  booktitle    = {Proceedings of the 2018 {USENIX} Annual Technical Conference, {USENIX}
                  {ATC} 2018, Boston, MA, USA, July 11-13, 2018},
  pages        = {387--400},
  publisher    = {{USENIX} Association},
  year         = {2018},
}

@article{Haimes1971_EpsilonConstraint,
  author  = {Haimes, Yacov Y. and Lasdon, Leon S. and Wismer, David A.},
  title   = {On a Bicriterion Formulation of the Problems of Integrated System Identification and System Optimization},
  journal = {IEEE Transactions on Systems, Man, and Cybernetics},
  volume  = {SMC-1},
  number  = {3},
  pages   = {296--297},
  year    = {1971},
  month   = jul,
  doi     = {10.1109/TSMC.1971.4308298}
}

@article{DBLP:journals/aim/Kotthoff14,
  author       = {Lars Kotthoff},
  title        = {Algorithm Selection for Combinatorial Search Problems: {A} Survey},
  journal      = {{AI} Mag.},
  volume       = {35},
  number       = {3},
  pages        = {48--60},
  year         = {2014},
  doi          = {10.1609/AIMAG.V35I3.2460},
}

@article{DBLP:journals/ac/Rice76,
  author       = {John R. Rice},
  title        = {The Algorithm Selection Problem},
  journal      = {Adv. Comput.},
  volume       = {15},
  pages        = {65--118},
  year         = {1976},
  doi          = {10.1016/S0065-2458(08)60520-3},
}

@article{Chaurasia2026IHFD,
  author  = {Chaurasia, B. and Verma, A. and Verma, P.},
  title   = {{IHFD}: An Improved Heartbeat-Based Failure Detector of {SVG} Class for Partially Synchronous Hierarchical Distributed Systems},
  journal = {Complex \& Intelligent Systems},
  year    = {2026},
  doi     = {10.1007/s40747-026-02423-z},
  url     = {https://doi.org/10.1007/s40747-026-02423-z}
}

@InProceedings{10.1007/failure-detectors-nn,
author="Lima, F{\'a}bio
and Mac{\^e}do, Raimundo",
title="Adapting Failure Detectors to Communication Network Load Fluctuations Using {SNMP} and Artificial Neural Nets",
booktitle="Dependable Computing",
year="2005",
publisher="Springer Berlin Heidelberg",
pages="191--205",
isbn="978-3-540-32092-0"
}

@inproceedings{DBLP:conf/usenix/RovelliCBPJE25,
  author       = {Davide Rovelli and
                  Pavel Chuprikov and
                  Philipp Berdesinski and
                  Ali Pahlevan and
                  Patrick Jahnke and
                  Patrick Eugster},
  title        = {Fi{D}e: Reliable and Fast Crash Failure Detection to Boost Datacenter
                  Coordination},
  booktitle    = {Proceedings of the 2025 {USENIX} Annual Technical Conference, {USENIX}
                  {ATC} 2025, Boston, MA, USA, July 7-9, 2025},
  pages        = {765--788},
  publisher    = {{USENIX} Association},
  year         = {2025},
}

@article{DBLP:journals/jacm/ChandraHT96,
  author       = {Tushar Deepak Chandra and
                  Vassos Hadzilacos and
                  Sam Toueg},
  title        = {The Weakest Failure Detector for Solving Consensus},
  journal      = {J. {ACM}},
  volume       = {43},
  number       = {4},
  pages        = {685--722},
  year         = {1996}
}

@article{IntroToFD/10.1145/1052796.1052806,
author = {Raynal, Michel},
title = {A short introduction to failure detectors for asynchronous distributed systems},
year = {2005},
issue_date = {March 2005},
publisher = {Association for Computing Machinery},
address = {New York, NY, USA},
volume = {36},
number = {1},
issn = {0163-5700},
IGNORE_doi = {10.1145/1052796.1052806},
journal = {SIGACT News},
month = mar,
pages = {53–70},
numpages = {18}
}

@article{StudyOnFD/Chaurasia2020,
author = {Chaurasia, Bhavana and Verma, Anshul},
year = {2020},
month = jul,
pages = {250-260},
title = {A Comprehensive Study on Failure Detectors of Distributed Systems},
volume = {64},
journal = {Journal of Scientific Research},
IGNORE_doi = {10.37398/JSR.2020.640235}
}

@article{PlanetLab/10.1145/956993.956995,
author = {Chun, Brent and Culler, David and Roscoe, Timothy and Bavier, Andy and Peterson, Larry and Wawrzoniak, Mike and Bowman, Mic},
title = {Planet{L}ab: an overlay testbed for broad-coverage services},
year = {2003},
issue_date = {July 2003},
publisher = {Association for Computing Machinery},
address = {New York, NY, USA},
volume = {33},
number = {3},
issn = {0146-4833},
IGNORE_doi = {10.1145/956993.956995},
journal = {SIGCOMM Comput. Commun. Rev.},
month = jul,
pages = {3–12},
numpages = {10}
}

@article{Yao2025/10.1145/3725536.3725549,
author = {Yao, Yuncheng},
title = {{MLFD}: The Implementation and Performance Evaluation of an {LSTM}-based, {S}mart{NIC}-Offloadable Failure Detector},
year = {2025},
issue_date = {March 2025},
publisher = {Association for Computing Machinery},
address = {New York, NY, USA},
volume = {52},
number = {4},
issn = {0163-5999},
IGNORE_doi = {10.1145/3725536.3725549},
journal = {SIGMETRICS Perform. Eval. Rev.},
month = mar,
pages = {25–27},
numpages = {3},
}

@inproceedings{Qi2023/10.1117/12.3004136,
author = {Wenbin Qi and Jingyi Zhao},
title = {{Machine learning applied to failure detection}},
volume = {12800},
booktitle = {Sixth International Conference on Computer Information Science and Application Technology (CISAT 2023)},
editor = {Huajun Dong and Shijie Jia},
organization = {International Society for Optics and Photonics},
publisher = {SPIE},
pages = {1280021},
year = {2023},
IGNORE_doi = {10.1117/12.3004136},
}

@article{DBLP:journals/corr/abs-2210-00134,
  author       = {Xiaonan Li and
                  Olivier Marin},
  title        = {Towards Implementing {ML}-Based Failure Detectors},
  journal      = {CoRR},
  volume       = {abs/2210.00134},
  year         = {2022}
}

@inproceedings{DBLP:conf/netys/BlanchardDBD14,
  author       = {Peva Blanchard and
                  Shlomi Dolev and
                  Joffroy Beauquier and
                  Sylvie Dela{\"{e}}t},
  title        = {Practically Self-stabilizing {P}axos Replicated State-Machine},
  booktitle    = {{NETYS}},
  series       = {Lecture Notes in Computer Science},
  pages        = {99--121},
  publisher    = {Springer},
  year         = {2014}
}

@article{DBLP:journals/jacm/DworkLS88,
  author       = {Cynthia Dwork and
                  Nancy A. Lynch and
                  Larry J. Stockmeyer},
  title        = {Consensus in the presence of partial synchrony},
  journal      = {J. {ACM}},
  volume       = {35},
  number       = {2},
  pages        = {288--323},
  year         = {1988}
}

@article{DBLP:journals/jacm/ChandraT96,
  author       = {Tushar Deepak Chandra and
                  Sam Toueg},
  title        = {Unreliable Failure Detectors for Reliable Distributed Systems},
  journal      = {J. {ACM}},
  volume       = {43},
  number       = {2},
  pages        = {225--267},
  year         = {1996}
}

@article{DBLP:journals/jacm/FischerLP85,
  author       = {Michael J. Fischer and
                  Nancy A. Lynch and
                  Mike Paterson},
  title        = {Impossibility of Distributed Consensus with One Faulty Process},
  journal      = {J. {ACM}},
  volume       = {32},
  number       = {2},
  pages        = {374--382},
  year         = {1985}
}

@article{DBLP:journals/jpdc/SensARM24,
  author       = {Pierre Sens and
                  Luciana Arantes and
                  Anubis Graciela de Moraes Rossetto and
                  Olivier Marin},
  title        = {{S}tab-{FD}: {A} cooperative and adaptive failure detector for wide area
                  networks},
  journal      = {J. Parallel Distributed Comput.},
  volume       = {186},
  pages        = {104803},
  year         = {2024}
}

@inproceedings{DBLP:conf/srds/DixitC10,
  author       = {M{\^{o}}nica Dixit and
                  Ant{\'{o}}nio Casimiro},
  title        = {{A}daptare-{FD}: {A} Dependability-Oriented Adaptive Failure Detector},
  booktitle    = {{SRDS}},
  pages        = {141--147},
  publisher    = {{IEEE} Computer Society},
  year         = {2010}
}

@inproceedings{DBLP:conf/ipps/TomsicSGAS15,
  author       = {Alejandro Z. Tomsic and
                  Pierre Sens and
                  Jo{\~{a}}o Garcia and
                  Luciana Arantes and
                  Julien Sopena},
  title        = {{2W}-{FD}: {A} Failure Detector Algorithm with {QoS}},
  booktitle    = {{IPDPS}},
  pages        = {885--893},
  publisher    = {{IEEE} Computer Society},
  year         = {2015}
}

@INPROCEEDINGS{S-AFD/Liu15,
  author={Liu, Jiaxi and Wu, Zhibo and Lan, Jinghui and Dong, Jian and Wu, Jin and Yu, Jiaxin},
  booktitle={2015 Fifth International Conference on Instrumentation and Measurement, Computer, Communication and Control (IMCCC)}, 
  title={A Low-Overhead Cooperative Failure Detector}, 
  year={2015},
  volume={},
  number={},
  pages={811-815},
  IGNORE_doi={10.1109/IMCCC.2015.177}
}

@article{DBLP:journals/sensors/DongRZL14,
  author       = {Jian Dong and
                  Xiao Ren and
                  Decheng Zuo and
                  Hongwei Liu},
  title        = {An Adaptive Failure Detector Based on Quality of Service in Peer-to-Peer
                  Networks},
  journal      = {Sensors},
  volume       = {14},
  number       = {9},
  pages        = {16617--16629},
  year         = {2014},
  IGNORE_doi          = {10.3390/S140916617},
}

@inproceedings{DBLP:conf/sac/SatzgerPTU07,
  author       = {Benjamin Satzger and
                  Andreas Pietzowski and
                  Wolfgang Trumler and
                  Theo Ungerer},
  title        = {A new adaptive accrual failure detector for dependable distributed
                  systems},
  booktitle    = {{SAC}},
  pages        = {551--555},
  publisher    = {{ACM}},
  year         = {2007}
}

@article{WeibullFD/journal.pone.0173666,
    IGNORE_doi = {10.1371/journal.pone.0173666},
    author = {Liu, Jiaxi AND Wu, Zhibo AND Wu, Jin AND Dong, Jian AND Zhao, Yao AND Wen, Dongxin},
    journal = {PLOS ONE},
    publisher = {Public Library of Science},
    title = {A {W}eibull distribution accrual failure detector for cloud computing},
    year = {2017},
    month = mar,
    volume = {12},
    pages = {1-16},
    number = {3},
}

@misc{EDFD/Xiong07,
author="Xiong, Naixue and D{\'{e}}fago, Xavier",
title="{ED} {FD}: Improving the {\(\Phi\)} accrual failure detector",
journal="Research report",
ISSN="0918-7553",
publisher="School of Information Science, Japan Advanced Institute of Science and Technology",
year="2007",
month=apr,
volume="IS-RR-2007-007",
pages="1-29",
URL="https://hdl.handle.net/10119/4799",
note="Technical Report IS-RR-2007-007, JAIST"
}

@inproceedings{DBLP:conf/srds/HayashibaraDYK04,
  author       = {Naohiro Hayashibara and
                  Xavier D{\'{e}}fago and
                  Rami Yared and
                  Takuya Katayama},
  title        = {The {\(\Phi\)} Accrual Failure Detector},
  booktitle    = {{SRDS}},
  pages        = {66--78},
  publisher    = {{IEEE} Computer Society},
  year         = {2004}
}

@article{2EFD/journal.pone.0191577,
    IGNORE_doi = {10.1371/journal.pone.0191577},
    author = {Liu, Jiaxi AND Wu, Zhibo AND Dong, Jian AND Wu, Jin AND Wen, Dongxin},
    journal = {PLOS ONE},
    publisher = {Public Library of Science},
    title = {An energy-efficient failure detector for vehicular cloud computing},
    year = {2018},
    month = jan,
    volume = {13},
    pages = {1-17},
    number = {1},
}

@inproceedings{DBLP:conf/dsn/BertierMS02,
  author       = {Marin Bertier and
                  Olivier Marin and
                  Pierre Sens},
  title        = {Implementation and Performance Evaluation of an Adaptable Failure
                  Detector},
  booktitle    = {{DSN}},
  pages        = {354--363},
  publisher    = {{IEEE} Computer Society},
  year         = {2002}
}

@article{DBLP:journals/tc/ChenTA02a,
  author       = {Wei Chen and
                  Sam Toueg and
                  Marcos Kawazoe Aguilera},
  title        = {On the Quality of Service of Failure Detectors},
  journal      = {{IEEE} Trans. Computers},
  volume       = {51},
  number       = {5},
  pages        = {561--580},
  year         = {2002}
}

\clearpage
\appendix
\section{Detector Details} \label{app:detector_details}

\subsection{Implementation Details}
\label{app:detector_implementation_details}

We implemented all failure detectors in Python (except as noted for Adaptare below), following the algorithms described in their original publications.
The ten offline detectors (\(\phi\)-FD, Chen, ED, Satzger, Stab, 2W, Bertier, 2E, WD, and BDBD) share a common interface and are evaluated in our trace-driven simulator: each detector is notified of every heartbeat in the recorded traces, with its sequence number and receipt time, and is periodically queried for its suspicion status.
The three real-time detectors (Adaptare, S-AFD, and Stab\(^C\)) run as asynchronous UDP processes on the deployment nodes.

\subparagraph*{Dependencies.}
The offline detectors use only the Python standard library and NumPy.
\(\phi\)-FD additionally uses SciPy for the normal distribution.
The real-time detectors use only the standard library.
For Adaptare, timeout bounds are computed with the authors' original implementation, which also depends only on the standard library.

\subparagraph*{Arrival history.}
Detectors that estimate arrival statistics over a sliding window use the \num{1000} most recent heartbeats of each monitored node.
Any exceptions follow the designs of the respective detectors.
2W combines this long window with a short window holding only the latest heartbeat.
2E keeps no window and instead smooths the observed delays exponentially (\(\alpha = 0.85\)).
Bertier estimates the expected arrival time over the window but derives its safety margin from a Jacobson-style exponentially smoothed estimator (\(\gamma = 0.1\)).
Adaptare follows the pull-style protocol of the original paper, with probe period \(\eta = T_{MR}(1 - C)\) for target mistake recurrence time \(T_{MR}\) and coverage \(C\), and computes its bound over the last \(h = 30\) round-trip times, the value used in the original evaluation.
S-AFD is round-based and keeps no arrival-time window.

\subsection{Detector Detail Tables} \label{app:detector_detail_table}

\Cref{tab:systematization_summary} complements \Cref{fig:fd_evolution_diagram} by summarizing the operational design dimensions used later to select evaluation modes and detector configurations.
\Cref{tab:detector_configs} complements \Cref{sec:evaluation} by detailing the configurations used in our evaluation for the detector implementations.

\begin{table}[ht]
\caption{Summary of the systematized failure detectors. B-AFD is systematized but not implemented; we evaluate instead its extension S-AFD (\Cref{sec:chronology}).}
\label{tab:systematization_summary}
\centering
\setlength{\tabcolsep}{3pt}
\begin{tblr}{
  width = \linewidth,
  colsep = 0.5em,
  rowsep = 0.15em,
  colspec = {
    @{}
    X[18,l]
    X[22,l]
    X[18,l]
    X[42,l]
    @{}
  },
  rows = {valign=m, font=\footnotesize},
  row{1} = {font=\bfseries\footnotesize},
  row{even} = {bg=gray!10},
}
\toprule
Algorithm
& Design dimension
& Detection/ output
& Main refinement \\
\midrule
Chen~\cite{DBLP:journals/tc/ChenTA02a}
& HB/timeout
& Push, binary
& QoS metrics; expected arrival; constant margin \\

Bertier~\cite{DBLP:conf/dsn/BertierMS02}
& HB/timeout
& Push, binary
& Jacobson-style dynamic margin \\

$\phi$-FD~\cite{DBLP:conf/srds/HayashibaraDYK04}
& Accrual
& Push, accrual
& Continuous suspicion level \\

ED~\cite{EDFD/Xiong07}
& Accrual
& Push, accrual
& Exponential inter-arrival model \\

Satzger~\cite{DBLP:conf/sac/SatzgerPTU07}
& Accrual
& Push, accrual
& Histogram/CDF suspicion estimate \\

Adaptare~\cite{DBLP:conf/srds/DixitC10}
& Probe with QoS
& Pull, binary
& Dependability-oriented parameter adaptation \\

BDBD~\cite{DBLP:conf/netys/BlanchardDBD14}
& Self-stabilizing
& Push, binary
& Bounded counters; no timestamps \\

B-AFD~\cite{DBLP:journals/sensors/DongRZL14}
& Probe with retransmission
& Pull, binary
& Retransmission-based QoS optimization \\

S-AFD~\cite{S-AFD/Liu15}
& Probe with cooperation
& Pull, binary
& Negative message sharing \\

2W~\cite{DBLP:conf/ipps/TomsicSGAS15}
& HB/timeout
& Push, binary
& Short and long windows \\

WD~\cite{WeibullFD/journal.pone.0173666}
& Accrual
& Push, accrual
& Weibull inter-arrival model \\

2E~\cite{2EFD/journal.pone.0191577}
& HB/timeout
& Push, binary
& last delay estimate; smoothed margin \\

Stab~\cite{DBLP:journals/jpdc/SensARM24}
& HB/timeout
& Push, binary
& Stability-calibrated margin \\

Stab\(^C\)~\cite{DBLP:journals/jpdc/SensARM24}
& HB/timeout with cooperation
& Push, binary
& Stability-calibrated margin; cooperative views \\
\bottomrule
\end{tblr}
\end{table}

\begin{table}[ht]
\caption{Parameter configurations used in the evaluation.
We vary parameters that primarily affect detection time across various ranges.
Notably, Adaptare and S-AFD do not expose an explicit parameter for controlling the message rate.
We therefore select different configurations that yield either \qty{10}{\Hz} or \qty{100}{\Hz} probing behavior.}
\label{tab:detector_configs}
\centering
\small
\begin{tblr}{
  colspec = {X[18,l,m] X[44,l] X[12,c] X[26,l]},
  colsep = 0.5em,
  rowsep = 0.15em,
  rows = {valign=m},
  row{1} = {font=\bfseries, halign=c},
  row{even} = {bg=gray!10},
}
\toprule
Algorithm & Parameter & No. of config. & Range of values \\
\midrule

Chen
& safety margin
& 27 
& \qtyrange{7}{12000}{\ms} \\

Bertier
& Jacobson-style variance multiplier
& 13
& \qtyrange{0.1}{30}{} \\

\(\phi\)-FD
& suspicion threshold
& 14
& \qtyrange{0.1}{16}{} \\

ED
& suspicion threshold
& 12
& \qtyrange{0.6}{0.99999}{} \\

Satzger
& suspicion threshold
& 12
& \qtyrange{0.6}{0.99999}{} \\

BDBD 
& safety margin coefficient
& 11
& \qtyrange{0}{5}{} \\

2W
& safety margin
& 27
& \qtyrange{7}{12000}{\ms} \\

WD
& suspicion threshold
& 12
& \qtyrange{0.6}{0.99999}{} \\

2E 
& safety margin coefficient \(\beta\)
& 26
& \qtyrange{0.01}{5000}{} \\

Stab
& initial safety margin
& 27
& \qtyrange{7}{12000}{\ms} \\

\midrule

Adaptare
& mistake-recurrence target \(T_{MR}\) 
& 12
& \qtyrange{1}{5000}{\second} \\

& coverage target (for \qty{10}{\Hz})
& 12
& \qtyrange{0.9}{0.99998}{}  \\

& coverage target (for \qty{100}{\Hz})
& 12
& \qtyrange{0.99}{0.999998}{}  \\

S-AFD 
& retry count per round \(rs\)
& 12
& \qtyrange{3}{100}{tries} \\

& round duration \(\Delta_s\) (for \qty{10}{\Hz})
& 12
& \qtyrange{0.3}{10}{\second} \\

& round duration \(\Delta_s\) (for \qty{100}{\Hz})
& 12
& \qtyrange{0.03}{1}{\second} \\

Stab\(^C\)
& initial safety margin
& 12
& \qtyrange{7}{3000}{\ms} \\
\bottomrule
\end{tblr}
\end{table}

\clearpage
\section{Supplementary Results for Heartbeat Rate and Host Load}\label{app:detailed_results}

\begin{figure}[ht]
    \centering
    \begin{subfigure}{\linewidth}
        \centering
        \includegraphics[width=\linewidth]{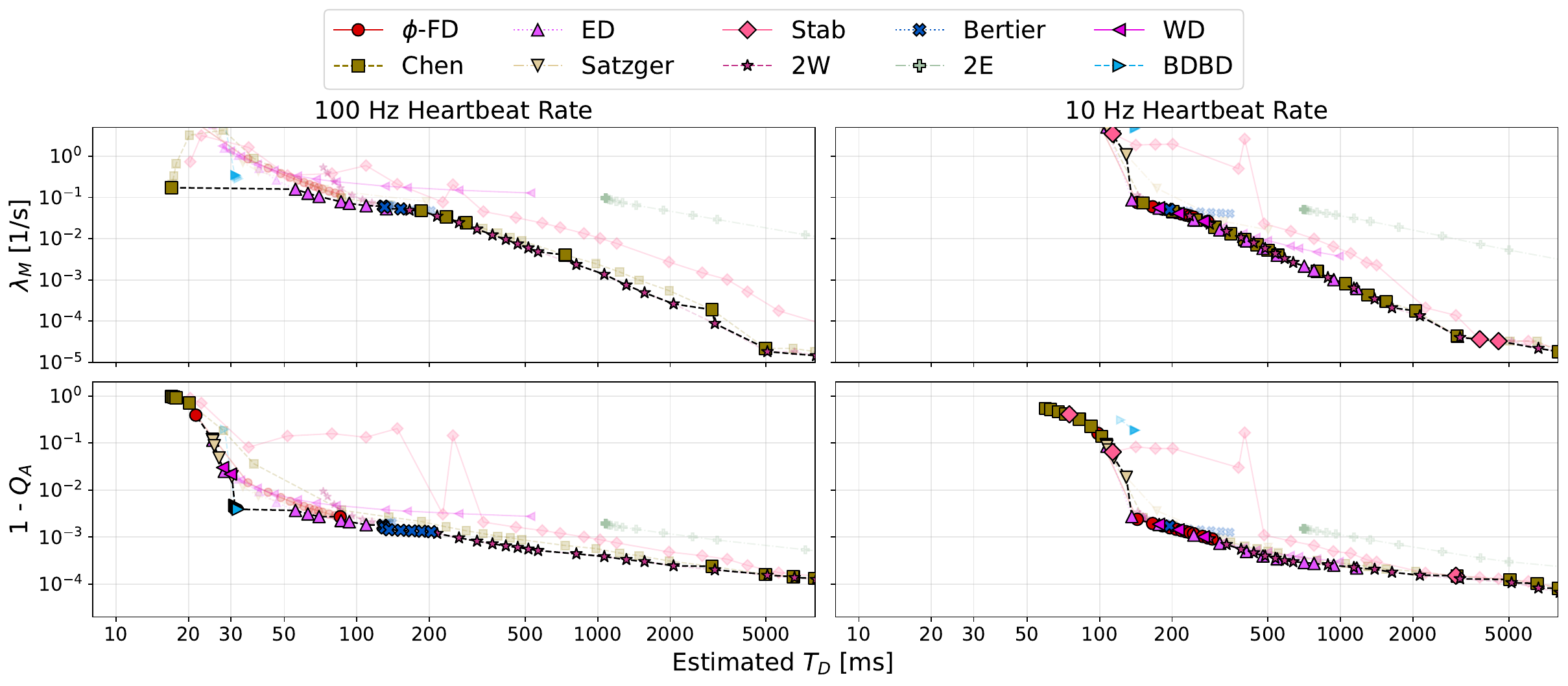}
        \caption{Results under the NC environment.}
        \label{fig:hb_green}
    \end{subfigure}
    
    \begin{subfigure}{\linewidth}
        \centering
        \includegraphics[width=\linewidth]{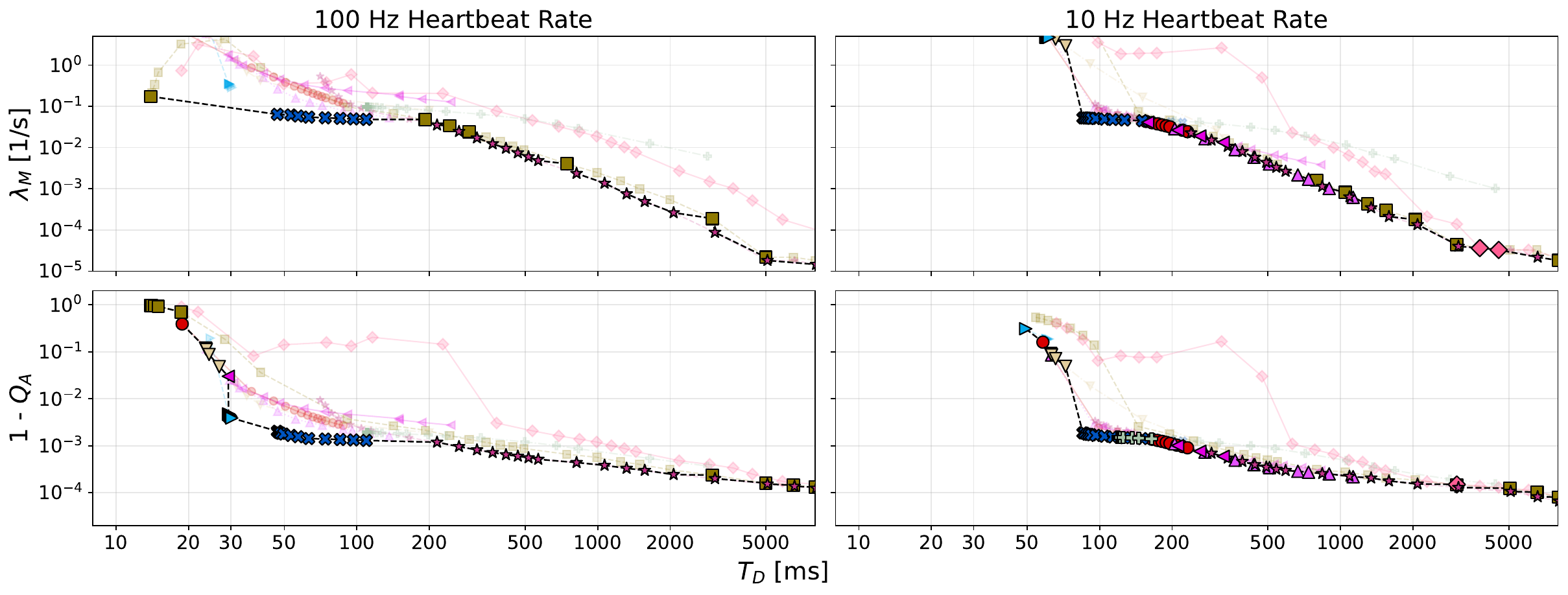}
        \caption{Results under the CI environment.}
        \label{fig:hb_red}
    \end{subfigure}

    \caption{Experiment 1: QoS metrics (\(\lambda_M\) and \(1 - Q_A\)) for offline detectors at \qty{100}{\Hz} and \qty{10}{\Hz} under (a) NC and (b) CI environments. All axes use log scale.}
    \label{fig:hb_100hz_vs_10hz}
\end{figure}

\begin{figure}[ht]
    \centering
    \begin{subfigure}{\linewidth}
        \centering
        \includegraphics[width=\linewidth]{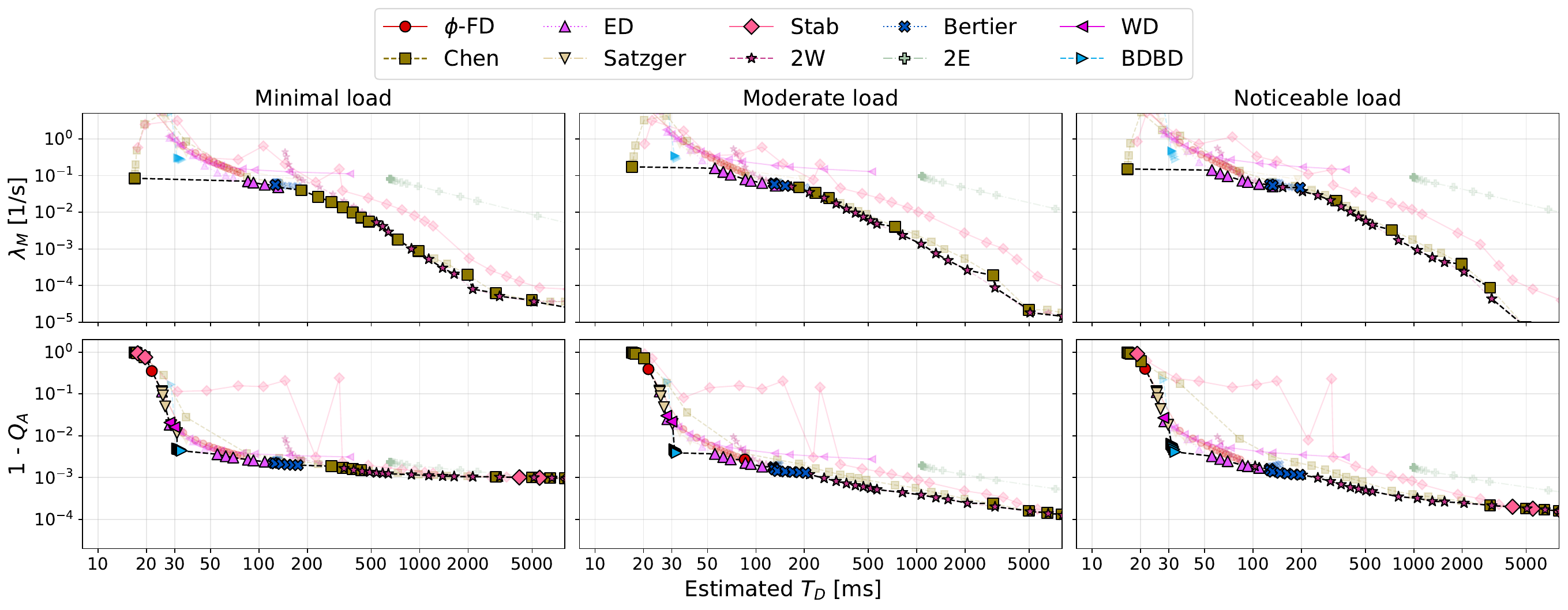}
        \caption{Results under the NC environment.}
        \label{fig:host_load_green}
    \end{subfigure}
    
    \begin{subfigure}{\linewidth}
        \centering
        \includegraphics[width=\linewidth]{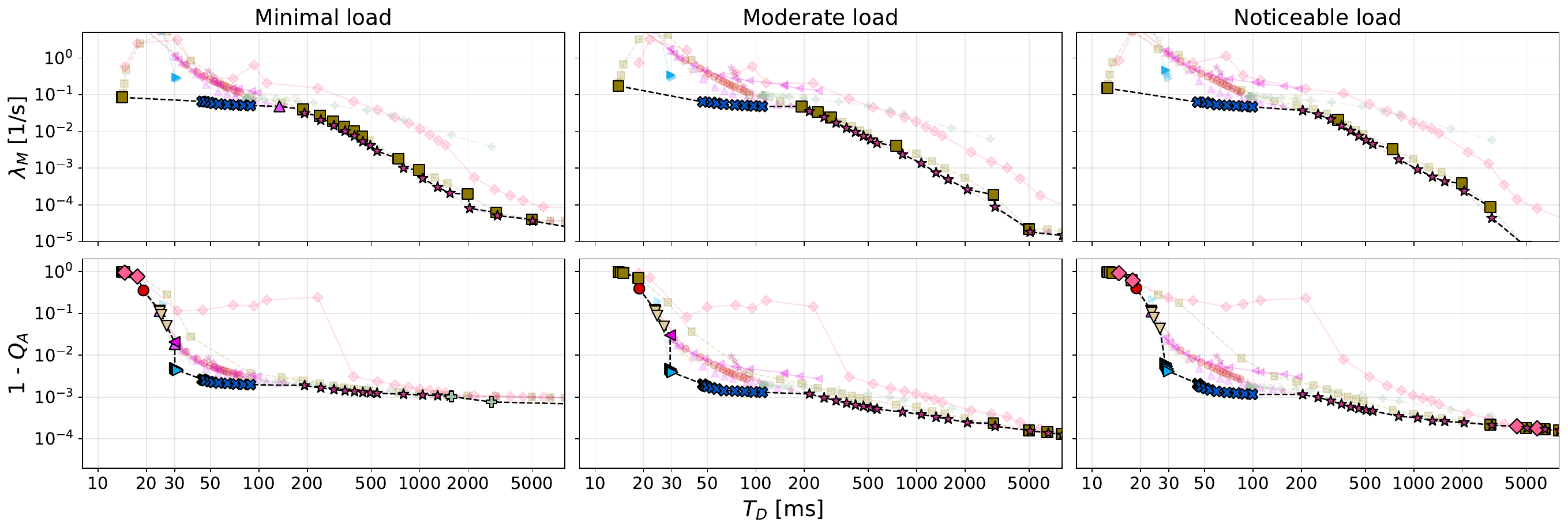}
        \caption{Results under the CI environment.}
        \label{fig:host_load_red}
    \end{subfigure}
   
    \caption{Experiment 2: QoS metrics (\(\lambda_M\) and \(1 - Q_A\)) for offline detectors under minimal, moderate, and noticeable host load in (a) NC and (b) CI environments. All axes use log scale.}
    \label{fig:host_load_comparisons}
\end{figure}

This appendix reports supplementary plots and tables for \Cref{sec:results}. 
These plots are provided for completeness and reproducibility, and they do not change the interpretation of the main results or the detector-selection guidance.
\Cref{fig:hb_100hz_vs_10hz} gives the full heartbeat-rate comparison, \Cref{fig:host_load_comparisons} gives the full host-load comparison, and \Cref{tab:load_best_results_app} presents all detection-time thresholds used to check the robustness of the reported trends.

\begin{table}[ht]
\caption{Lowest mistake rates \(\lambda_M\) and query error probabilities \(1 - Q_A\) achieved under each host-load setting, crash environment, and detection-time threshold.}
\label{tab:load_best_results_app}
\centering
\begin{tblr}{
  colspec = {l c c c c  c c c},
  colsep = 0.5em,
  rowsep = 0.15em,
  row{odd} = {bg=gray!10},
  row{1-2} = {font=\bfseries, halign=center, bg=white},
}
\toprule
\SetCell[r=2]{c} Load & \SetCell[r=2]{c} Env.  & \SetCell[r=2]{c} Metric & \SetCell[c=5]{c} Detection Time Threshold & & & & \\
\cline{4-8} 
& & & $T_D {\le} 50$ & $T_D {\le} 100$ & $T_D {\le} 200$ & $T_D {\le} 500$ & $T_D {\le} 2000$ \\
\midrule
min. & CI & $\lambda_M$ & 0.062 & 0.051 & 0.031 &  0.0041  & \num{2.0e-4} \\
mod. & CI & $\lambda_M$ & 0.064 & 0.049 & 0.048 & 0.0074  & \num{4.9e-4} \\
not. & CI & $\lambda_M$ & 0.063 & 0.047 & 0.047 &  0.0075  & \num{3.9e-4} \\
min. & CI & $1 - Q_A$ & 0.0024 & 0.0020  & 0.0019 & 0.0013  & 0.0010  \\
mod. & CI & $1 - Q_A$ & 0.0018 & 0.0013 & 0.0013 & \num{6.0e-4}  & \num{3.0e-4}  \\
not. & CI & $1 - Q_A$ & 0.0017 & 0.0012  & 0.0012 & \num{5.4e-4} & \num{2.6e-4}   \\

\midrule
min. & NC & $\lambda_M$ & 0.084 & 0.064 & 0.040 &  0.0056  & \num{2.0e-4} \\
mod. & NC & $\lambda_M$ & 0.173 & 0.072 & 0.048 & 0.0074  & \num{4.9e-4} \\
not. & NC & $\lambda_M$ & 0.151 & 0.067 & 0.047 &  0.0075  & \num{3.9e-4} \\
min. & NC & $1 - Q_A$ & 0.0044 & 0.0026  & 0.0020 & 0.0014  & 0.0011  \\
mod. & NC & $1 - Q_A$ & 0.0039 & 0.0021  & 0.0013 & \num{6.0e-4}  & \num{3.0e-4}  \\
not. & NC & $1 - Q_A$ & 0.0041 & 0.0019  & 0.0012 & \num{5.4e-4} & \num{2.6e-4}   \\

\bottomrule
\end{tblr}
\end{table}

\end{document}